\documentclass{aastex631}

\usepackage{amsmath}
\usepackage{graphicx} % 用于插入图片
\usepackage{booktabs}
\usepackage{multirow}
\usepackage{xcolor}
\usepackage{fancyhdr}
\usepackage{hyperref}
\usepackage{ulem}

\begin{document}
\pagestyle{fancy}
\fancyhf{}

\rfoot{\thepage}
\title{Detection of four cold Jupiters through combined analyses of radial velocity and astrometry data\footnote{This paper includes data gathered with the 6.5 meter Magellan Telescopes located at Las Campanas Observatory, Chile.}}

\author{Yiyun Wu}
\affiliation{State Key Laboratory of Dark Matter Physics, Tsung-Dao Lee Institute \& School of Physics and Astronomy, Shanghai Jiao Tong University, Shanghai 201210, China}

\author{Guang-Yao Xiao}
\affiliation{State Key Laboratory of Dark Matter Physics, Tsung-Dao Lee Institute \& School of Physics and Astronomy, Shanghai Jiao Tong University, Shanghai 201210, China}

\author{R. Paul Butler}
\affiliation{Earth and Planets Laboratory, Carnegie Institution for Science, 5241 Broad Branch Road, NW, Washington, DC 20015, USA}

\author{Fabo Feng}
\altaffiliation{ffeng@sjtu.edu.cn}
\affiliation{State Key Laboratory of Dark Matter Physics, Tsung-Dao Lee Institute \& School of Physics and Astronomy, Shanghai Jiao Tong University, Shanghai 201210, China}

\author{Stephen A. Shectman}

\affiliation{Observatories of the Carnegie Institution for Science, 813 Santa Barbara Street, Pasadena, CA 91101, USA}

\author{Jeffrey D. Crane}
\affiliation{Observatories of the Carnegie Institution for Science, 813 Santa Barbara Street, Pasadena, CA 91101, USA}

\author{Johanna K. Teske}
\affiliation{Earth and Planets Laboratory, Carnegie Institution for Science, 5241 Broad Branch Road, NW, Washington, DC 20015, USA}

%\author{C. G. Tinney}
%\affiliation{School of Physics and Australian Centre for Astrobiology, University of New South Wales, Sydney 2052, Australia}
\author{Sharon X. Wang}
\affiliation{Department of Astronomy, Tsinghua University, Beijing 100084, People’s Republic of China}

\author{Yuri Beletsky}
\affiliation{Las Campanas Observatory, Carnegie Institution of Washington, Colina el Pino, Casilla 601, La Serena, Chile}

\author{Jennifer A. Burt}
\affiliation{Jet Propulsion Laboratory, California Institute of Technology, 4800 Oak Grove drive, Pasadena CA 91109, USA}

\author{Tansu Daylan}
\affiliation{Department of Physics and McDonnell Center for the Space Sciences, Washington University, St. Louis, MO 63130, USA}

\author{Matias Diaz}
\affiliation{Las Campanas Observatory, Carnegie Institution of Washington, Colina El Pino S/N, La Serena, Chile}

\author{Diana Dragomir}
\affiliation{Department of Physics and Astronomy, University of New Mexico, 1919 Lomas Boulevard NE, Albuquerque, NM 87131, USA}

\author{Erin Flowers}
\affiliation{Department of Astrophysical Sciences, Princeton University, 4 Ivy Lane, Princeton, NJ 08540, USA}

\author{Sydney Jenkins}
\affiliation{Department of Physics, Massachusetts Institute of Technology, 77 Massachusetts Avenue, Cambridge, MA 02139, USA}
\affiliation{Kavli Institute for Astrophysics and Space Research, Massachusetts Institute of Technology, Cambridge, MA 02139, USA}

\author{Shubham Kanodia}
\affiliation{Earth and Planets Laboratory, Carnegie Science, 5241 Broad Branch Road, NW, Washington, DC 20015, USA}

\author{Sangeetha Nandakumar}
\affiliation{INCT, Universidad de Atacama, calle Copayapu 485, Copiapó, Atacama, Chile}

%\author[orcid=0000-0002-7670-670X,sname='Rice']{Malena Rice}
\author{Malena Rice}
\affiliation{Department of Astronomy, Yale University, 219 Prospect Street, New Haven, CT 06511, USA}
%\email{malena.rice@yale.edu}

\author{Avi Shporer}
\affiliation{Kavli Institute for Astrophysics and Space Research, Massachusetts Institute of Technology, Cambridge, MA 02139, USA}

%\author{Ian Thompson}
%\affiliation{Observatories of the Carnegie Institution for Science, 813 Santa Barbara Street, Pasadena, CA 91101, USA}

\author{Sam Yee}
\affiliation{Center for Astrophysics | Harvard \& Smithsonian, 60 Garden Street, Cambridge, MA 02138, USA}

\author{George Zhou}
\affiliation{University of Southern Queensland, Centre for Astrophysics, West Street, Toowoomba, QLD 4350, Australia}

\author{Zitao Lin}
\affiliation{Department of Astronomy, Tsinghua University, Beijing 100084, People’s Republic of China}

\begin{abstract}

Cold Jupiters play a crucial role in planet formation and dynamical evolution. Since their initial discovery around 47 UMa, they have attracted significant interest, yet their formation mechanisms remain uncertain, underscoring the need to expand the known population.

In this work, we combine RV data with Gaia astrometry using Hipparcos–Gaia proper-motion anomalies over a 25-year baseline. By jointly modeling both datasets with the MCMC framework, we constrain planetary masses, orbital inclinations, and three-dimensional orbital architectures. This reduces RV degeneracies and improves mass determinations. Four cold Jupiters are reported: HD 68475 b and HD 100508 b are each the first confirmed planet in their systems, with orbital periods ${7832}_{-323}^{+463}$ d and $5681\pm42$ d and dynamical masses of $5.16^{+0.53}_{-0.47}, M_{\text{Jup}}$ and $1.2^{+0.30}_{-0.18}, M{\text{Jup}}$, respectively. In multi-planet systems, HD 48265 c has a period of ${10418}_{-1400}^{+2400}$ d and a mass of $3.71^{+0.68}_{-0.43}, M_{\text{Jup}}$, while HD 114386 c orbits at ${444.00}_{-0.88}^{+0.93}$ d with a minimum mass of $0.37^{+0.03}_{-0.03}, M_{\text{Jup}}$.
 
The two planets in the HD 48265 system may exhibit a significant mutual inclination, making it a target for testing the von-Zeipel–Kozai–Lidov mechanism. HD 68475 b is a promising candidate for future direct imaging with ELT/METIS. We identified a Jupiter analog with the longest known orbital period among planets with masses between 0.5 and 2 $M_{\text{Jup}}$, implying that a substantial population of cold Jupiters likely awaits discovery by Gaia.

This study expands the sample of cold Jupiters with constrained orbits and dynamical masses, demonstrating the value of combining radial velocity and astrometry in exoplanet research.

\end{abstract}

\section{Introduction}\label{sec:intro}
Since the confirmation of the first exoplanet in 1992 (\citealt{1992Natur.355..145W}) and the discovery of the hot Jupiter 51 Pegasi b (\citealt{1995Natur.378..355M}), exoplanetary research has unveiled a diverse array of planetary systems, including the first cold Jupiter 47 UMa (\citealt{1996ApJ...464L.153B}) and eccentric planet 70 Vir (\citealt{1996ApJ...464L.147M}). Among them, giant planets — especially cold Jupiters — are of particular interest. Generally, cold Jupiters are defined as giant planets with semi-major axes greater than 1 au and masses exceeding $0.3\,M_{\mathrm{Jup}}$. According to core accretion theory, giant planet formation typically occurs beyond the snow line, where solid material can efficiently accumulate to form massive cores that subsequently accrete gas envelopes \citep{1996Icar..124...62P, 2004ApJ...616..567I}. However, the precise formation mechanisms of cold Jupiters remain uncertain, with open questions regarding their core masses, migration histories, and overall occurrence rates across different stellar environments. 
Beyond the questions surrounding their own formation, the influence of cold Jupiters on inner planetary systems is also a subject of debate. 
Some observational studies have reported a high co-occurrence rate between cold Jupiters and inner super-Earths \citep{2018AJ....156...92Z, 2019AJ....157...52B}, while others suggest that cold Jupiters may suppress inner planet formation \citep{2023A&A...677A..33B}. Recent work by \citet{2024AJ....168..268V}, \citet{2025arXiv250520035B} and \citet{2025ApJ...982L...7B} investigated the relationship between cold Jupiters and small planets. While no significant correlation is observed for most systems, \citet{2025ApJ...982L...7B} reported that a strong positive correlation emerges for K-type and higher-mass stars with [Fe/H]$>$0.1, suggesting that the connection may depend on stellar mass and metallicity. Overall, the influence of cold Jupiters on small planets, such as super-Earths, remains unresolved.
Moreover, understanding how cold Jupiters influence planetary system dynamics — via phenomena such as orbital misalignment
\citep{2015ARA&A..53..409W, 2022PASP..134h2001A}, and mechanisms including secular interactions such as the von-Zeipel–Kozai–Lidov cycles (or ZKL cycles; \citealt{von_Zeipel1910, 1962AJ.....67R.579K, 1962P&SS....9..719L, 2016ARA&A..54..441N}) and planet–planet scattering \citep{2008ApJ...686..580C, 2008ApJ...686..603J} — requires precise knowledge of their true masses and three-dimensional orbits. 
Therefore, expanding the sample of well-characterized cold Jupiters is essential for testing formation and migration theories, reducing uncertainties in their impact on small planets, and evaluating their dynamical effects on planetary architectures.

The radial velocity method (RV) measures the Doppler shift of stellar spectral lines and reveals the variation of a star’s velocity along the line of sight of the observer over time. The precision of stellar Doppler measurements improved by two orders of magnitude from 1980 to 1995, from 300 $\rm m~s^{-1}$ to 3 $\rm m~s^{-1}$ \citep{1996ApJ...464L.153B}. This led to the discovery of the first several hundred exoplanets, including hot Jupiters, eccentric planets, cold Jupiters, Saturn-mass planets, Neptune-mass planets, the first super-Earths, and helped  to establish the broader field of exoplanets. Over the past 30 years the improvement in Doppler precision has slowed, from 3 m s$^{-1}$ to $\sim$1 m s$^{-1}$ \citep{2016PASP..128f6001F}, representing an increase in precision of roughly two orders of magnitude during the initial 15 years, followed by an additional factor of approximately 10 over the subsequent three decades. With modern instruments such as ESPRESSO, NEID, and KPF, single-measurement precision is now reaching 30 $\sim$ 50 cm s$^{-1}$ \citep{2021A&A...645A..96P,2016SPIE.9908E..7HS,2024SPIE13096E..09G}.
 However, the radial velocity technique has inherent limitations, such as providing only a minimum mass estimate ($ M_{\rm p}\sin i$) and being susceptible to false positives induced by stellar jitter—caused by oscillations, spots, and magnetic cycles,  thus requiring complementary methods.

For long period planets, it is possible to combine Doppler velocities with precision space-based astrometry. This also yields the final orbital element, inclination, revealing the full 3-D orbit and the true planet mass. Several groups have developed techniques that combine Hipparcos and Gaia astrometry
\citep{2018NatAs...2..883S, 2018ApJS..239...31B,2019MNRAS.490.5002F, 2019A&A...623A..72K,2020MNRAS.497.2096X, 2024AJ....167..250V}. These approaches can break the degeneracy between mass and inclination and recover the three-dimensional reflex motion of host stars, especially by using proper-motion anomalies between Hipparcos and Gaia. However, most previous works have only used a single Gaia data release, which may limit the reliability for planets with orbital periods comparable to the satellite baseline. Moreover, proper-motion anomaly methods alone suffer from an inclination degeneracy, making it difficult to distinguish between prograde and retrograde orbits—though this limitation can be overcome by combining multiple Gaia data releases (e.g., GDR2 and GDR3; \citealt{GaiaCollaboration2018, GaiaCollaboration2023}).

In this study, the joint astrometric–radial velocity method developed by \cite{2022ApJS..262...21F} and modified by \cite{2023MNRAS.525..607F} is adopted. Gaia epoch astrometry is simulated using the Gaia Observation Forecast Tool (GOST\,\footnote{\url{https://gaia.esac.esa.int/gost/}}), and a linear astrometric model is applied to recover the nonlinear stellar motion. 
Our simulations explicitly account for observational gaps as dead points, addressing the potential impact of GOST's $\sim80\%$ prediction accuracy. Previous studies (e.g., \citealt{2023MNRAS.525..607F}) show that random variations in observation times do not significantly affect orbital precision, ensuring that our results are not biased by these gaps.
With this approach, the orbital inclination and true planetary mass can be better constrained. The method has been successfully applied to nearby cold Jupiters, including $\varepsilon\,$Ind A b, $\varepsilon\,$Eridani b \citep{2023MNRAS.525..607F, 2025MNRAS.539.3180F}, and HD 222237 \citep{2024MNRAS.534.2858X}.  

In this paper, the RV data are drawn from high-precision survey programs including CORALIE (\citealt{2001Msngr.105....1Q}), HARPS (\citealt{2000SPIE.4008..582P}), MIKE (\citealt{2003SPIE.4841.1694B}), and PFS (\citealt{2006SPIE.6269E..31C, 2008SPIE.7014E..79C, 2010SPIE.7735E..53C}). 
We combine these RVs with high-precision astrometry from Gaia (\citealt{2017A&A...601A..19G}), where recent advances have significantly enhanced this combined approach, providing valuable constraints on the masses and orbital properties of cold Jupiters.

The paper is structured as follows. Section \ref{sec:data} describes the observational datasets, while Section \ref{sec:fit} details the multi-method fitting approach. In Section \ref{sec:result}, we present the key results. Section \ref{sec:discussion} discusses the potential for direct imaging observations of certain signals and the formation mechanisms of cold Jupiters.

\section{Data} \label{sec:data}
We make use of both radial velocity and astrometric measurements from multiple instruments and missions to constrain the orbits of long-period exoplanets. To account for the intrinsic zero-point differences between different instruments, we treat the RV offsets as free parameters during the orbital fitting process. In particular, we distinguish between HARPSpre and HARPSpost to reflect the instrumental fiber upgrade 2018, treating them as separate datasets with individual zero-point offsets. The radial velocity curves are then fitted by combining data from these various instruments.

 %\subsection{CORALIE}
 CORALIE (Contrast Radial velocity List Expeditor, \citealt{2001Msngr.105....1Q}) is one of the early spectrographs for exoplanet detection, installed on the 1.2 - meter telescope at the La Silla Observatory of the European Southern Observatory (ESO). It has a resolution of approximately 70,000, covering a wavelength range of about 390 to 680 nm.

 %\subsection{HARPS}
 HARPS (High Accuracy Radial velocity Planet Searcher, \citealt{2000SPIE.4008..582P}) began operation in 2003. It has a spectral resolution of about 120,000 and covers a wavelength range of approximately 380-  690 nm used to derive the radial velocities.
 It has contributed to numerous exoplanet discoveries, such as HD 41004 B b, which orbits one component of a binary star system \citep{2003A&A...404..775Z}, and the super-Earth GJ 1214 b, a warm sub-Neptune  \citep{2014Natur.505...69K}.

%\subsection{MIKE}
 MIKE (Magellan Inamori Kyocera Echelle,  \citealt{2003SPIE.4841.1694B}) is installed on one of the 6.5-meter Magellan telescopes at the Las Campanas Observatory in Chile.  It has been in operation since 2004 and features high stability, with a blue arm covering 320–500 nm ($R\approx 28,000- 83,000$) and a red arm covering 490–1000 nm ($R \approx 22,000- 65,000$).
 
 \subsection{PFS}
 PFS (Carnegie Planet Finder Spectrograph; \citealt{2010SPIE.7735E..53C}) is a spectrograph installed on the 6.5‑meter Magellan II (Clay) telescope at Las Campanas Observatory in Chile. 
Prior to 2018, it provided a resolving power of $R\sim80,000$ ; since then, the resolution has increased to $\sim130,000$.
 It covers a wavelength range of 388 to 668 nm and is known for its high precision and stability.
To precisely measure the features in the stellar spectrum, PFS uses an iodine cell as a wavelength reference, which allows accurate determination of the positions of spectral lines due to planetary perturbations, corresponding to variations in the star's radial velocity as observed. The PFS played a key role in confirming multiple Earth-mass planet candidates around the M dwarf GJ 667C \citep{2012ApJ...751L..16A}, demonstrating its critical contribution to the detection and characterization of dynamically packed, potentially habitable planetary systems around low-mass stars.

\subsection{Hipparcos and Gaia}
 While radial velocity observations provide critical line-of-sight motion data, astrometric measurements offer complementary information in the plane of the sky. This allows for a more complete, three-dimensional reconstruction of the star's orbit and can remove the $\sin i$ degeneracy that limits RV-only detections to providing only a minimum mass measurement.
 One source of astrometric data is from Hipparcos, an astronomical satellite launched by the European Space Agency (ESA) in 1989. Using a 0.45-meter telescope, it provided precise measurements of positions and proper motions for approximately 118,000 stars \citep{1997ESASP.402....1P}. Gaia \citep{2017A&A...601A..19G}, launched by ESA in 2013, is a successor to Hipparcos with similar goals and improved technology. It employs a dual-telescope design consisting of two rectangular primary mirrors, each measuring 1.45~m~$\times$~0.5~m, enabling micro-arcsecond astrometry for over a billion stars.

Techniques that combine Hipparcos and Gaia astrometry have recently been applied to measure mutual inclinations and long-term dynamical architectures of planetary systems (e.g., \citealt{2025arXiv250220561T, 2025AJ....169..200Z, 2021AJ....162...12V, 2025AJ....169...22A}). These studies demonstrate that joint astrometric solutions can reveal relative orbital orientations between inner and outer companions, offering key insights into the dynamical histories of multi-planet systems.

  In this study, we use astrometric data from both the Hipparcos and Gaia catalogs, including stellar parallaxes ($\varpi$), positions (right ascension $\alpha$ and declination $\delta$), and proper motions (in R.A.: $\mu_{\alpha}$ and in Decl.: $\mu_{\delta}$). Specifically, we incorporate Hipparcos intermediate astrometry data (IAD; \citealt{2007A&A...474..653V}) and Gaia DR2 and DR3 astrometry \citep{2018A&A...616A...1G, 2023A&A...674A...1G}. Since Gaia IAD is not publicly available, we use the GOST to obtain the epochs and scan-angle geometry of Gaia observations, which are combined with our astrometric model to simulate along-scan measurements. These simulated data, along with catalog astrometry and radial velocity observations, are jointly modeled while accounting for zero-point offsets and jitter for each instrument. The long baseline ($\sim$25 yr) between Hipparcos and Gaia enables tighter constraints on the three-dimensional motion of host stars, helping to break the $M_{\rm p} \sin i$ degeneracy and improve orbital solutions for long-period exoplanets.

\subsection{Stellar parameters}
In addition to observational data, stellar parameters are crucial for determining planetary parameters. The candidate host stars span spectral types from G5 to K3, covering evolutionary stages from the main sequence to subgiants. %HD 48265 (G5 IV/V) exhibits characteristics of a star transitioning off the main sequence, while HD 100508 (K1 IV) is already in the subgiant phase. In contrast, HD 68475 (K2 V) and HD 114386 (K3 V) are typical K-type dwarfs.
These stars have masses ranging from 0.80 to 1.38 $M_\odot$, relatively low activity levels, and effective temperatures between 4800 K and 5700 K. Their relatively stable and slowly evolving stellar environments are favorable for the long-term dynamical evolution of cold giant planets, under conditions that share key similarities with those of the outer Solar System. The detailed stellar parameters are summarized in Table~\ref{table: stellarparameter}. The stellar masses of HD 48265 and HD 114386 were derived from atmospheric parameters \citep{2021A&A...656A..53S} and fitted using isochrones from \citep{2005A&A...436..127J}, yielding more precise values with smaller errors compared to TIC.

To address the potential influence of stellar activity, we present two analyses: the Generalized Lomb--Scargle (GLS) periodogram (Figure~\ref{fig: gls}) and the Pearson correlation plot (Figure~\ref{fig: pearson}). The GLS periodogram compares the radial velocity (RV) signal with stellar activity indicators, including the Mount Wilson S-index, H$\alpha$, and the window function (WF), showing no significant peaks at the planetary periods. The Pearson correlation, with a coefficient of $|r| < 0.3$ between RV and the S-index, indicates a very weak linear relationship. Both analyses suggest that the observed planetary signals are not related to stellar activity.

\begin{table*}[!htb] 
%\begin{table*}[]
\hspace*{-2cm}
\centering
\caption{Stellar parameters of host stars of the studied planetary systems.\label{table: stellarparameter}}

\renewcommand{\arraystretch}{1.5}

\resizebox{1\textwidth}{!}{

\begin{tabular}{ccccccccccccc} 
\toprule 
\hline
Name & HIP ID& spectral type &$V$ & $B-V$ & $T_{\text{eff}}$ & $\log g $ & $[\text{Fe/H}]$ &$M_\star$ & Parallax&$\chi^2$& HGCA significance&Reference\\

 & & &(mag)&(mag) & (K) & $\log[\mathrm{cm s^{-2}}]$ & (dex) & ($M_{\odot}$)& (mas)& &$\sigma$& \\
\hline
HD 48265&31895& G5IV/V& 8.03& 0.75&$5805\pm 34$&$4.06\pm 0.07$&$0.39\pm0.03$ &$1.38\pm 0.06$&$11.01\pm0.02$&6.77&1.83&1,3,4,7\\ 
HD 68475&40051&K2VC&8.78&0.91&$5022\pm 113$&$4.560 \pm 0.08$&$0.07\pm0.09$&$0.83\pm0.10$&$29.99\pm0.01$&48.77&6.57&1,2,3,6\\
HD 100508&56363 &K1IV&7.74&0.83&$5419\pm 112$&$4.39\pm0.08$&$0.33\pm0.03$&$0.94\pm0.12$ &$30.51\pm0.16$&12.70&2.92&1,2,3\\
HD 114386&64295&K3VC&8.71&1.09&$4895\pm62$&$4.28\pm 0.13$&$-0.10\pm0.03 $&$0.80\pm0.03$&$35.74\pm0.02$& 3.95&1.08&1,3,5,7\\
\hline
\bottomrule 
\end{tabular}}
%\vspace{0.8em}
\begin{flushleft}
\textbf{References:}  
[1] \citet{2000A&A...355L..27H} – B and V magnitudes.  
[2] \citet{2006AJ....132..161G} – Spectral types for HD 68475 and HD 100508.  
[3] \citet{2018AJ....156..183S,2019AJ....158..138S} – Effective temperature, Star mass, [Fe/H], and $\log g$ from TIC. 
[4] \citet{1978mcts.book.....H} – Spectral type of HD 48265.  
[5] \citet{1972AJ.....77..486U} – Spectral type of HD 114386.  
[6] \citet{2008A&A...485..571J} – [Fe/H] of HD 68475.
[7]\citet{2021A&A...656A..53S} - [Fe/H], $\log g $, Effective temperature of HD 48265 and HD 114386.  
[8]\citet{2020yCat.1350....0G} - Parallax.
[9]\citet{2021ApJS..254...42B} - The $\chi^2$ values from the HGCA catalog for each star, as well as the corresponding significance expressed in units of $\sigma$.
\end{flushleft}

\end{table*}

\section{Method} \label{sec:fit}

RV measurements capture the component of a star's motion along the line of sight to the observer on Earth. Using radial velocity measurements of the host star, a planet’s orbital period, minimum mass, RV semi-amplitude, and orbital eccentricity can be determined.
Astrometry, on the other hand, measures the tangential component of a star's motion. By combining these two methods, the orbital inclination can be decoupled from the planetary mass, allowing for the determination of the planet's true mass.
 The methods employed in this work are discussed in detail in \cite{2023MNRAS.525..607F} and \cite{2024MNRAS.534.2858X}.

 Given the 25-year observation baseline between Hipparcos and Gaia, the position and proper motion changes of the primary star during this period may indicate perturbations caused by unseen celestial bodies (e.g., giant planets, brown dwarfs). These data provide a non-linear tangential motion of the star in the celestial projection. Combined with the line-of-sight motion observed through radial velocity, it allows us to fully determine the star's Keplerian motion in three-dimensional space and construct the 3D orbit of the planetary system, which is vital for understanding planet formation and evolution.
 
 When fitting orbits, we aim to find a set of parameter solutions that make the theoretical values predicted by the RV and astrometry models match the observed data as closely as possible. To efficiently explore the parameter space, we use the Parallel Tempering Markov Chain Monte Carlo (\texttt{ptemcee}, \citealt{2016MNRAS.455.1919V}) algorithm, a variant of MCMC that runs multiple chains at different temperatures to enhance sampling efficiency and avoid local maxima. This approach is particularly effective for the multimodal and correlated posteriors often seen in orbital parameter estimation.
 For an ideal two-body system, we use the aforementioned MCMC algorithm to sample the basic parameters: orbital period $P$, RV semi-amplitude $K$, eccentricity $e$, argument of periastron $\omega$ of stellar reflex motion, orbital inclination $i$, longitude of ascending node $\Omega$, mean anomaly $M_{0}$ at the minimum epoch of RV data and five astrometric offsets ($\Delta \alpha*$, $\Delta \delta$, $\Delta \varpi$, $\Delta \mu_{\alpha*}$ and $\Delta \mu_\delta$) of barycenter relative to GDR3. The semi-major axis $a$ of the planet relative to the host, the mass of planet $M_{\rm p}$, and the epoch of periastron passage $T_{\rm p}$ can be derived from above orbital elements.
 In particular, the stellar mass will be set as a Gaussian prior in our fitting.

For the RV-only fit, the likelihood function is employed as follows:
\begin{equation}
    \begin{aligned}
    \mathcal{L}_{\mathrm{RV}}=\prod_{j=1}^{N_{\mathrm{RV}}}\prod_{k=1}^{N_{\mathrm{inst}}}\frac{1}{\sqrt{2\pi(\sigma_{j,k}^2+\sigma_{\mathrm{jit},k}^2)}}\exp\left(-\frac{(\nu_{j,k}-\hat{\nu}_{j,k}-\gamma_k)^2}{2(\sigma_{j,k}^2+\sigma_{\mathrm{jit},k}^2)}\right),
    \end{aligned}
\end{equation}
%where $N_{\text{RV}}$ is the number of RV measurements. $\sigma_k$ is the individual measurement uncertainty, and $J_{\text{RV}}$ is the jitter term.
where \( N_{\mathrm{RV}} \) and \( N_{\mathrm{inst}} \) represent the total number of RV measurements and the number of instruments, respectively. \( \gamma_k \) denotes the RV offset for the \( k \)-th instrument, and \( \sigma_{\mathrm{jit},k} \) represents the additional jitter term to account for unmodeled stellar or instrumental noise. 
The formal uncertainty of each measurement is denoted as \( \sigma_{j,k} \).
The model-predicted radial velocity at epoch \( t_j \), denoted as \( \hat{v}_{j,k} \), is calculated as $\hat{v}_j = K \left[ \cos(\omega + \nu(t_j)) + e \cos(\omega) \right]$ where $\nu(t_j)$ is the true anomaly.

The likelihood function for the Gaia astrometry is defined as:

\begin{equation}
    \begin{aligned}
        \mathcal{L}_{\text{Gaia}} = \prod_{j=1}^{N_{\text{DR}}} (2\pi)^{-5/2} \left( \det \Sigma_j' \right)^{-1/2} \exp \left[ -\frac{1}{2} \left( \hat{\vec{\Delta \iota}}_j - {\Delta \vec{\iota}}_j \right)^T \left( \Sigma_j \left(1 + J_j\right) \right)^{-1} \left( \hat{\vec{\Delta \iota}}_j - {\Delta \vec{\iota}}_j \right) \right],
    \end{aligned}
\end{equation}
where \( N_{\text{DR}} \) is the number of Gaia data releases used in the analyses. To avoid numerical errors, the catalog astrometry at each epoch \( t_j \) is defined relative to the Gaia DR3 reference epoch by subtracting the Gaia DR3 parameters from the corresponding values in the input catalog (either Gaia DR2 or DR3), i.e.,
${\Delta \vec{\iota}}_j \equiv \left( \Delta\alpha_{*j}, \Delta\delta_j, \Delta\varpi_j, \Delta\mu_{\alpha j}, \Delta\mu_{\delta j} \right) = \left( (\alpha_j - \alpha_{\text{DR3}})\cos\delta_j, \delta_j - \delta_{\text{DR3}}, \varpi_j - \varpi_{\text{DR3}}, \mu_{\alpha j} - \mu_{\alpha j, \text{DR3}}, \mu_{\delta j} - \mu_{\delta j, \text{DR3}} \right).$
The fitted astrometry for epoch \( t_j \) is denoted as \( \hat{\vec{\Delta \iota}}_j \). 
\( \Sigma_j (1 + J_j) \) represents the jitter-corrected covariance matrix for the five-parameter solutions of Gaia DRs, where \( j = 1 \) and \( 2 \) represent Gaia DR2 and DR3, respectively. Usually, only DR2 and DR3 are used, and thus \( N_{\text{DR}} = 2 \).

The abscissae of Hipparcos $\hat{\xi}$ are modeled by adding the reflex motion onto the linear model, and the likelihood is calculated as follows:
%hip likelihood
\begin{equation}
    \begin{aligned}
        \mathcal{L}_{\mathrm{hip}} = \prod^{N_{\text{epoch}}}_{j} [2\pi(\sigma_j^2 + J^2_{\text{hip}})]^{-1/2} \exp\left[-\frac{(\hat{\xi_j} - \xi_j)^2}{2(\sigma_j^2 + J^2_{\text{hip}})}\right]
    \end{aligned}
\end{equation}
where  \( N_{\text{epoch}} \) is the number of  epochs of Hipparcos IAD.  $\sigma_j$ is the individual measurement uncertainty, and $J_{\text{hip}}$ is the jitter term.

We derive the likelihoods for the RV, Gaia, and Hipparcos data through the above; the total likelihood is obtained as follows:

\begin{equation}
    \begin{aligned}
        \mathcal{L}=   \mathcal{L}_{\mathrm{RV}}\cdot \mathcal{L}_{\mathrm{hip}}\cdot \mathcal{L}_{\mathrm{gaia}}
    \end{aligned}
\end{equation}

In systems with multiple companions, mutual gravitational interactions between planets are neglected. This approximation is quantitatively justified using the Laplace--Lagrange estimates for the nodal and pericenter precession rates. For HD~48265, the inferred secular timescale is $\tau_{\mathrm{sec}} \sim 6\times10^{4}\,\mathrm{yr}$, implying an interaction-induced RV drift of only $\sim 0.1\,\mathrm{m\,s^{-1}}$ over a 20-year baseline---well below the typical 1--3\,m\,s$^{-1}$ uncertainties. HD~114386 shows similarly negligible interaction amplitudes. Therefore, the host-star motion can be accurately modeled as the superposition of independent Keplerian signals from each companion.

%This approximation is quantitatively justified using the Laplace–Lagrange estimate for the nodal and pericenter precession rates, indicating that these interactions are much smaller than the dominant star–planet interactions in our target systems \citep{1999ssd..book.....M}. With this approach, the motion of the host star is treated as the superposition of independent Keplerian signals from each companion.

\section{Results} \label{sec:result}

This work presents planetary systems identified primarily using long-term radial velocity data from the PFS survey, along with complementary archival observations
 from other RV facilities.
 Combined with astrometric data from Gaia and Hipparcos, we selected four systems from the initial PFS sample — HD 48265, HD 68475, HD 100508, and HD 114386.  These systems show relatively significant signals and benefit from the availability of additional RV and astrometric data, making them well-suited for comprehensive orbital modeling and characterization. Their planetary masses and orbital parameters are further analyzed in detail.

\subsection{HD 48265}

HD 48265 is a G5 IV star located 87.4 pc from the Sun \citep{1997ESASP.402....1P}, with a metallicity of [Fe/H]$=0.39\pm 0.03$\,dex \citep{2021A&A...656A..53S}. Through the analysis of 4.4 yr of observations with the MIKE echelle spectrograph, \cite{2009ApJ...693.1424M} confirmed the orbital parameters of the inner planet  HD 48265 b ($P_{\rm b}$ = $762\pm50$ d, $e_{\rm b} = 0.24\pm0.10$ and $M_{\rm p}\sin{i}=1.2\,M_{\rm Jup}$). \cite{2017MNRAS.466..443J} refined and improved these parameters using new HARPS and CORALIE data. \cite{2018A&A...615A.175B} found good agreement with previous studies regarding $M_{\rm b}\sin{i}$ and period, except for eccentricity, with $M_{b}\sin i=1.53^{+0.05}_{-0.05}$, $P_{b}$ = 778.51$^{+5.38}_{-5.18}$ d and $e_{b} = {0.21}_{-0.10}^{+0.09}$. A two-planet solution for the system was sought using residual periodicity and an initial zero eccentricity setting for an additional planet, but satisfactory convergence was not achieved.

In this paper, we report the discovery of a long-period outer cold Jupiter, HD 48265 c, with a period of $P_{c}={28.5}_{-3.9}^{+6.7}$ yr. By combining radial velocity and astrometric data, we determine the true mass of the companion to be \(M_\mathrm{c} =4.09_{-0.20}^{+0.24}\, M_\mathrm{Jup}\). The planet exhibits an orbital eccentricity of \(e_c\ = 0.42^{+0.08}_{-0.06}\), and a semi-major axis of \(a_c = {10.4}_{-1.0}^{+1.6}\, \mathrm{au}\). The orbital inclination is constrained to \(i_c = 89^{+29}_{-29}\) deg, and the longitude of the ascending node is \(\Omega_c = 196^{+29}_{-29}\) deg.  The values of \(i_c\) and \(\Omega_c\) are derived using the third convention (or astrometric convention) as defined in \cite{2019MNRAS.490.5002F}. 
Nevertheless, as shown in Table~\ref{table: stellarparameter}, the HGCA astrometric acceleration for HD~48265 corresponds to a significance of 1.8$\sigma$, and therefore the derived astrometric constraints should be interpreted with appropriate caution.

The inner signal, HD 48265 b's parameters are be updated to: $e_b=0.35\pm0.02,\ P_b=789.6\pm1.1 \,\mathrm{d},\,  a={1.87}_{-0.04}^{+0.02}\,\mathrm{au}$. The minimum mass has been derived to be $M_\mathrm{b}\sin i =1.67_{-0.06}^{+0.06}\, M_\mathrm{Jup}$.

\begin{figure}[htp]
    \centering
    \includegraphics[width=12cm]{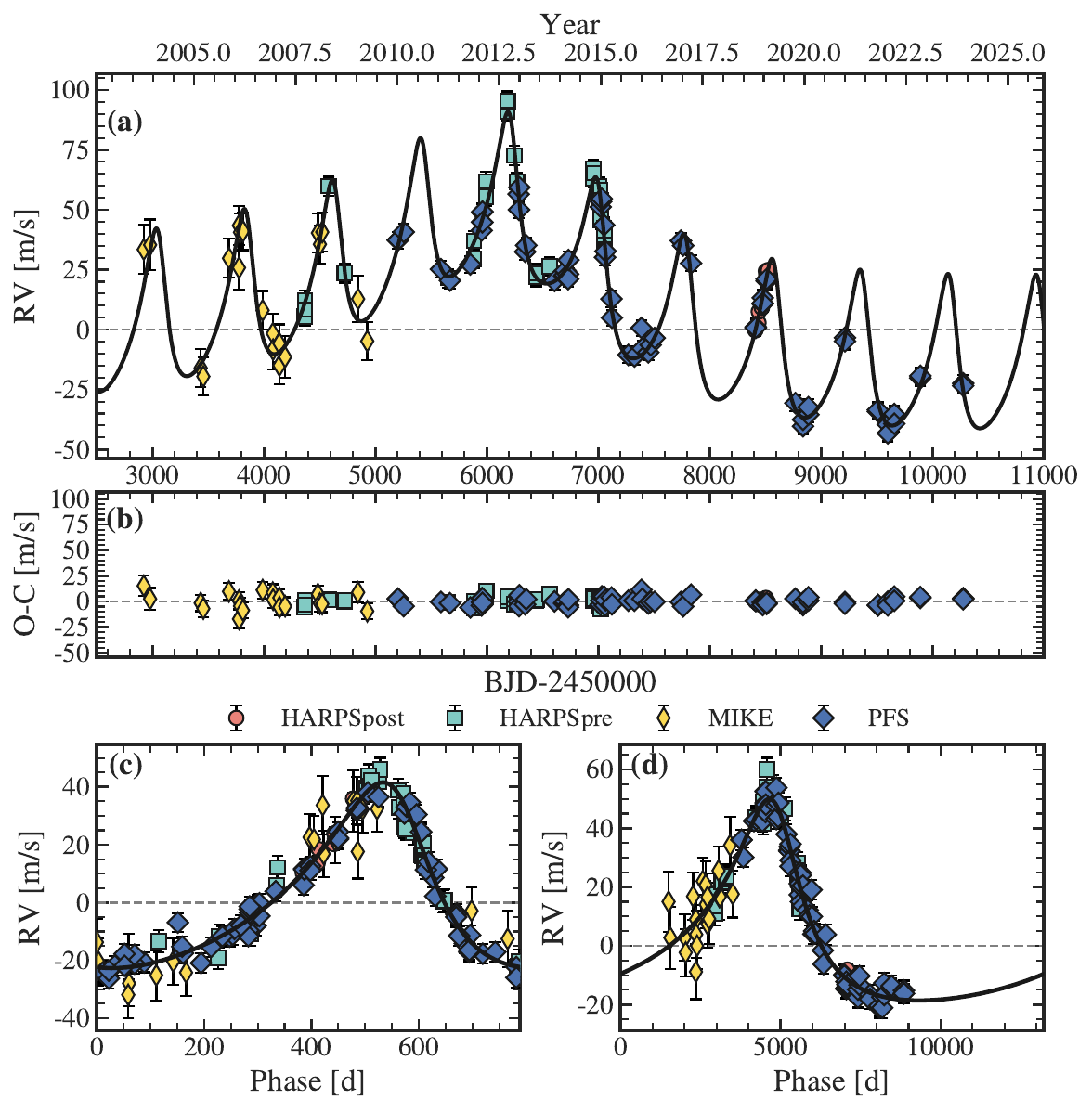}
    \caption{(a) RV curve of HD 48265. The points with error bars denote the RV measurements and associated uncertainties. The black line denotes the best-fit orbit. (b) Residual (O-C) between observation and model. Panels (c) and (d) show the RV signals from each of the two planets separately, with all data folded to the respective orbital period of each planet.
    \label{fig: HD48265RV}}
\end{figure}

 When the astrometric signal of the inner planet is included in the global fit, the best-fit orbital solution yields two symmetric families for the inclination: \( i_b = 11.7^{+11.0}_{-3.4} \, \mathrm{deg} \) and \( i_b = 160.1^{+8.3}_{-31.0} \, \mathrm{deg} \). The corresponding longitudes of the ascending node are \( \Omega_b = 40^{+19}_{-15} \, \mathrm{deg} \) for the prograde solution and \( \Omega_b = 80^{+26}_{-27} \, \mathrm{deg} \) for the retrograde one. The planet’s mass is estimated to be \( M_b = {7.4}_{-3.9}^{+3.7}\, M_\mathrm{Jup} \). These two solutions arise because the orbital period of this planet is much shorter than the Hipparcos mission baseline, making its astrometric signature effectively undetectable in the Hipparcos data. Consequently, the orbit is constrained only by two Gaia epochs, which are insufficient to break the degeneracy between prograde and retrograde configurations \citep{2025MNRAS.539.3180F}.Additional astrometric measurements from future Gaia data releases or other facilities will be essential to resolve this ambiguity.
 Figure ~\ref{fig: HD48265RV} presents the radial velocity (RV) analysis for HD 48265, with astrometric fitting provided in Figure ~\ref{fig: ast}. Astrometric data from Gaia and model predictions for HD 48265 are compared in Figure ~\ref{fig: HD48265box}, where we highlight the differences between the catalog measurements and our model’s predictions.

\subsection{HD 68475}

HD 68475  is a K2 V main-sequence star with an effective temperature of $5022 \pm 113 $ K, a visual magnitude of V = 8.78 mag, and a color index of $B\!-\!V$ = 0.910 mag \citep{2000A&A...355L..27H,2006AJ....132..161G}. We find this star to host a long-period cold Jupiter, HD 68475 b, with an orbital period of $P_b=21.44^{+1.27}_{-0.88}$ yr, a semi-major axis of $a_b=7.27^{+0.39}_{-0.36}\,\mathrm{au}$ , and an eccentricity of $e_b=0.62^{+0.02}_{-0.02}$. Its eccentricity is relatively high compared to the known population of long-period cold Jupiters. 

\begin{figure}[htp]
    \centering
    \includegraphics[width=12cm]{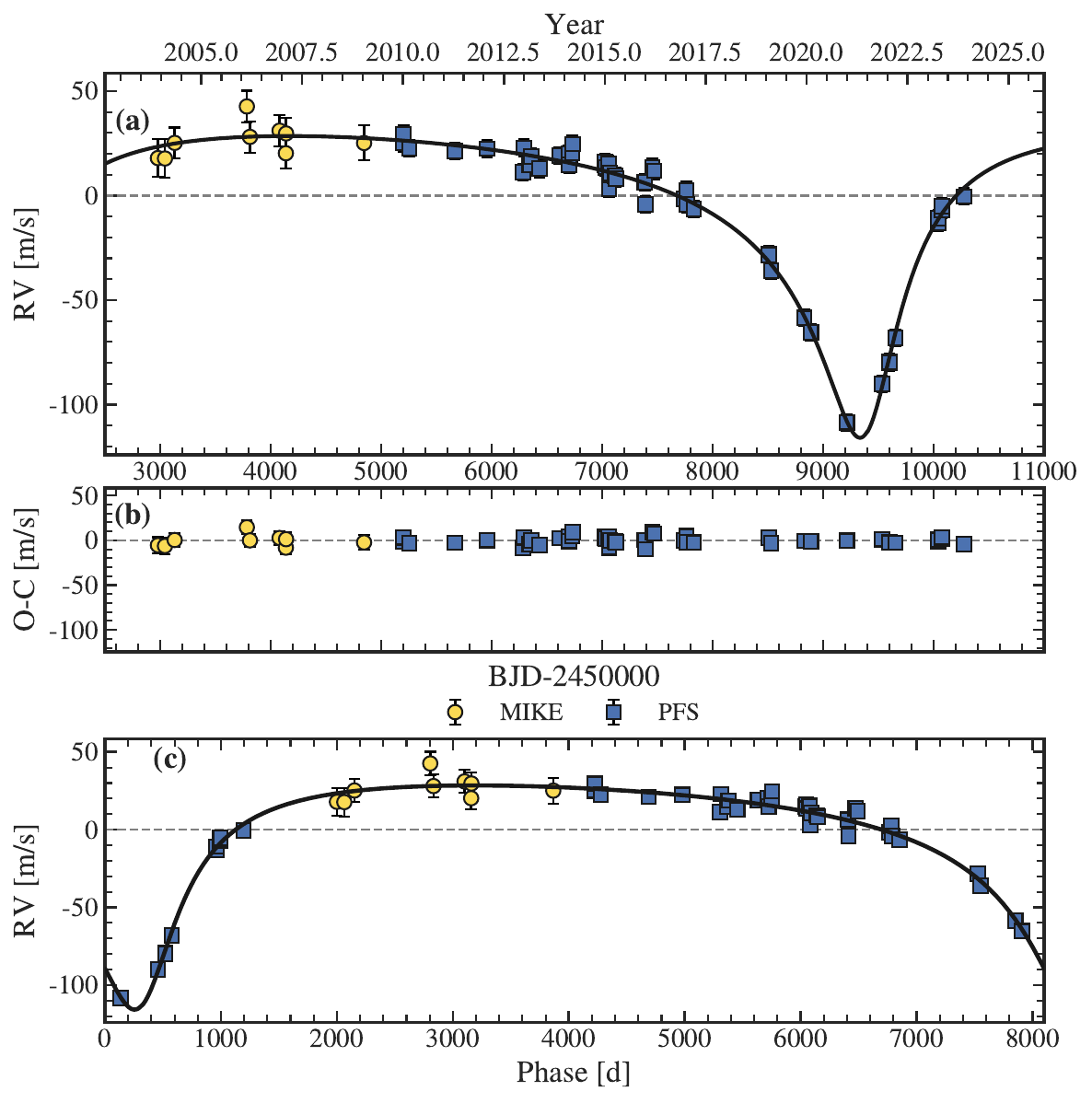}
        \caption{RV curve of HD 68475. The symbols are the same as Figure ~\ref{fig: HD48265RV}.}
    \label{fig: HD68475RV}
\end{figure}

\begin{figure}[htp]
    \centering
 \includegraphics[width=12cm]{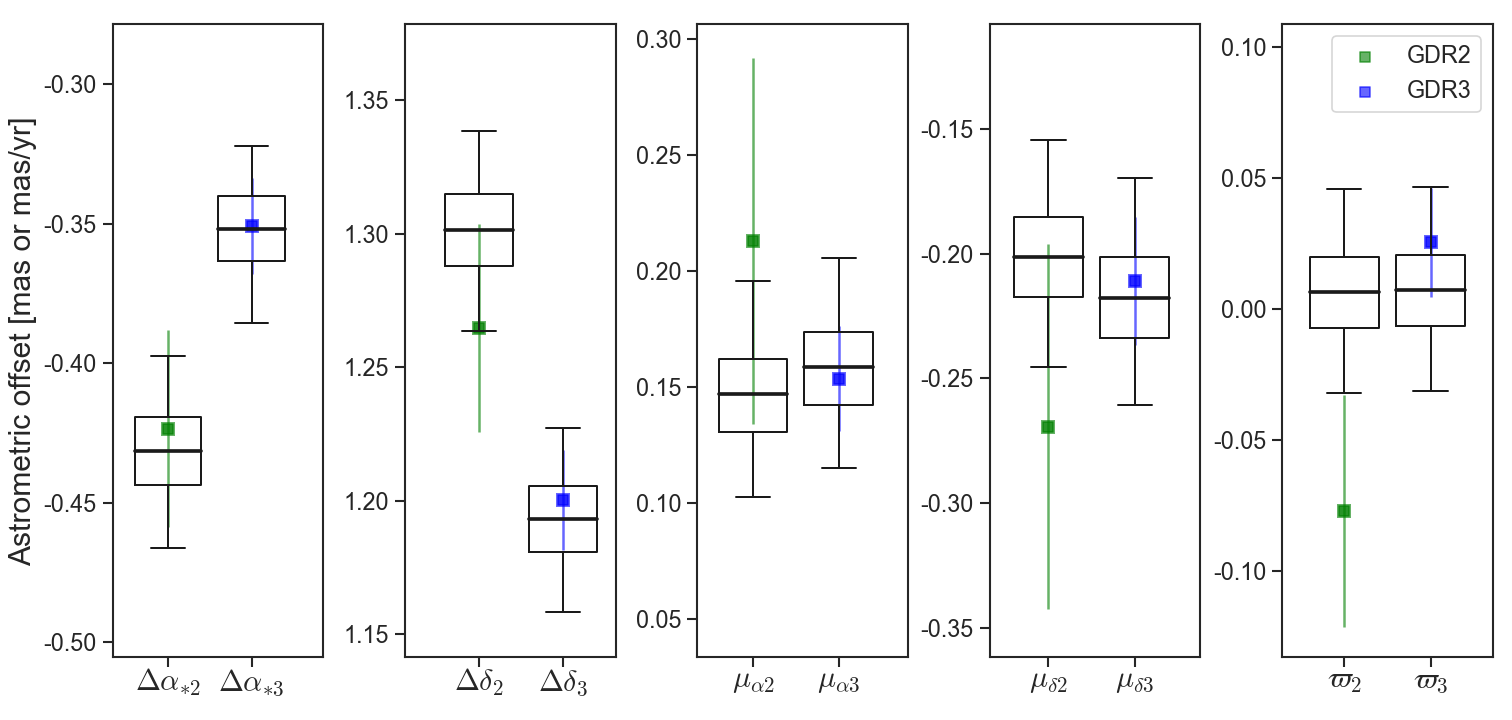}
        \caption{Comparing the five-parameter astrometry of the model prediction to GDR2 and GDR3 astrometry. The barycentric motion of the HD 68475 system has been subtracted for both the catalog Gaia data (square) and the predictions (boxplot). The inner thick line, edge of the box, and whisker, respectively, denote the median, $1 \sigma$ uncertainty and $3 
         \sigma$ uncertainty. The Gaia astrometric uncertainty has been adjusted for the error inflation factor. The subscripts of the label of the x-axis correspond to the Gaia release number.}
    \label{fig: HD68475box}
\end{figure}

\begin{figure}[htp]
    \centering
 \includegraphics[width=12cm]{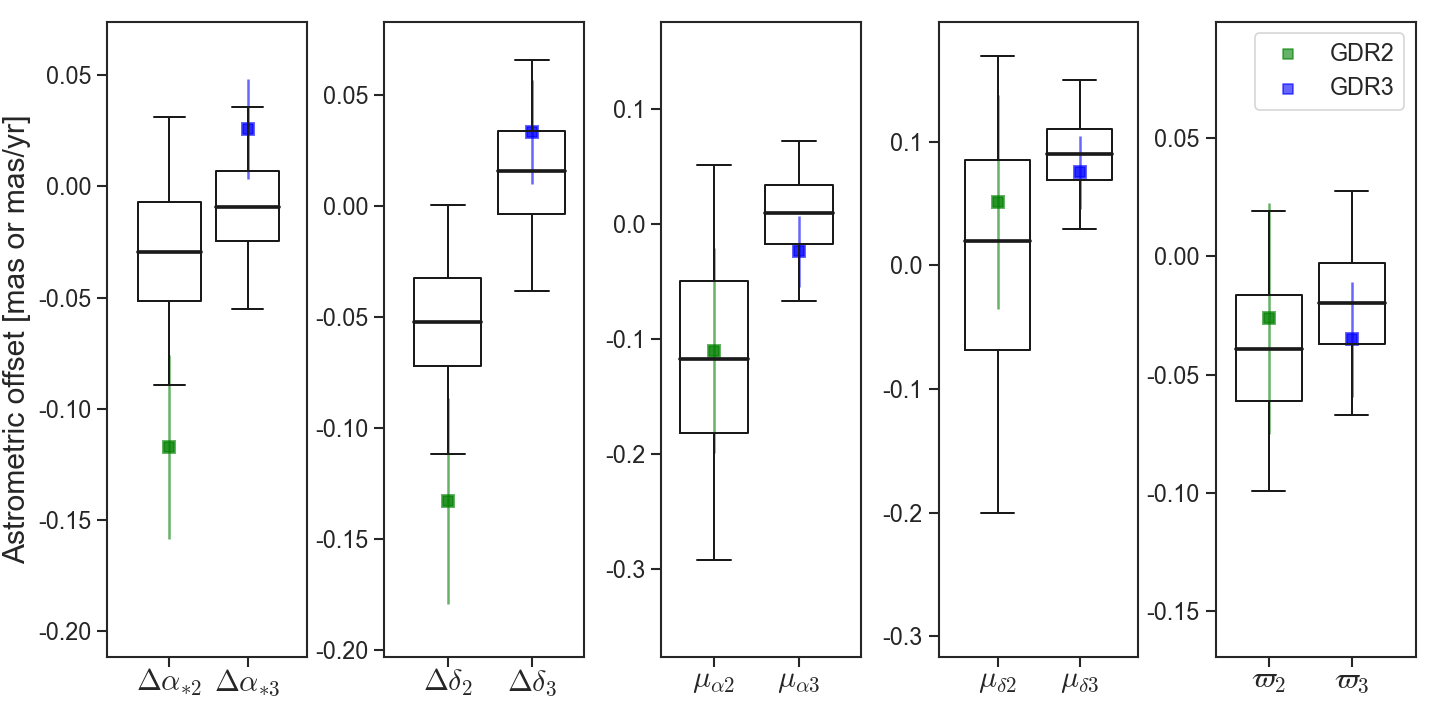}
        \caption{Boxplot for HD 48265. The symbols are the same as Figure ~\ref{fig: HD68475box}.}
    \label{fig: HD48265box}
\end{figure}

Using high-precision RV data from MIKE and PFS, combined with astrometry measurements, the orbital inclination is constrained to \( i_b = 87.66^{+21.77}_{-30.36} \,\text{deg} \).
 This enables the determination of the planet's mass as $M_\mathrm{b}=5.16^{+0.53}_{-0.47} \,M_\mathrm{{Jup}}$ and the longitude of the ascending node as $\Omega_b=327.2^{+8.4}_{-8.4}\,\text{deg}$. Located at a distance of $33.34\,$pc from the Sun, and given the moderate separation between the planet and its host star, this planetary companion presents a promising target for direct imaging with next-generation ground-based telescopes. 
 Figure~\ref{fig: HD68475RV} presents the radial velocity (RV) data for HD 68475, while Figure~\ref{fig: ast} illustrates its corresponding astrometric fitting. The comparison of Gaia astrometric data with the model predictions for HD 68475 is shown in Figure~\ref{fig: HD68475box}, demonstrating the discrepancies between the catalog values and our model.

\subsection{HD 114386}
HD 114386 is a K3 V star with V = 8.73 mag, $B\!-\!V$ = 0.982 mag, and an effective temperature of $4895\pm62$ K, $\log  g $ = $4.28\pm0.13$ cgs, [Fe/H]= $-0.10\pm0.03$, and a mass of $M_\star=0.80\pm 0.03\,M_{\odot}$ \citep{2000A&A...355L..27H,2019AJ....158..138S,2018AJ....156..183S,1972AJ.....77..486U,2021A&A...656A..53S}. The discovery of HD 114386 b was reported by \cite{2004A&A...415..391M}, and its orbital parameters were determined to be: $ P_b=937.7\pm15.6 \,\mathrm{d}$, $e_b=0.23\pm0.03$,  $M_b\sin i=1.24 \, M_\mathrm{Jup}$, and semi-major axis $ a_b=1.65 \,\mathrm{au}$. Subsequently, \cite{2017AJ....153..136S} updated these parameters to $M_b\sin i =1.14\pm0.13\,M_\mathrm{Jup}$, $e_b=0.23\pm0.03$, and $a_b=1.73\pm0.03 \,\mathrm{au}$.
 
While analyzing the RV data of this planetary system, we detected a potential signal of an additional inner planet with an orbital period of $P = 1.21\, \mathrm{yr}$ (approximately 444 d), consistent with the indication reported by \citep{2011arXiv1109.2497M}. Motivated by this, we adopted a two-planet model for the system. As a result, the eccentricity of HD 114386 b decreases to $e_b = 0.02^{+0.01}_{-0.01}$, significantly lower than the previously reported value of $e_b = 0.23$. The orbital inclination of planet b is constrained to $i_b = 57^{+24}_{-16}\,\mathrm{deg}$, and its mass is calculated to $1.44^{+0.36}_{-0.23}\, M_\mathrm{Jup}$. Other derived parameters include $a_b = 1.86^{+0.07}_{-0.08}\, \mathrm{au}$ and $\Omega_b = 278^{+19}_{-24}\, \mathrm{deg}$.

\begin{figure}[htp]
    \centering
    \includegraphics[width=12cm]{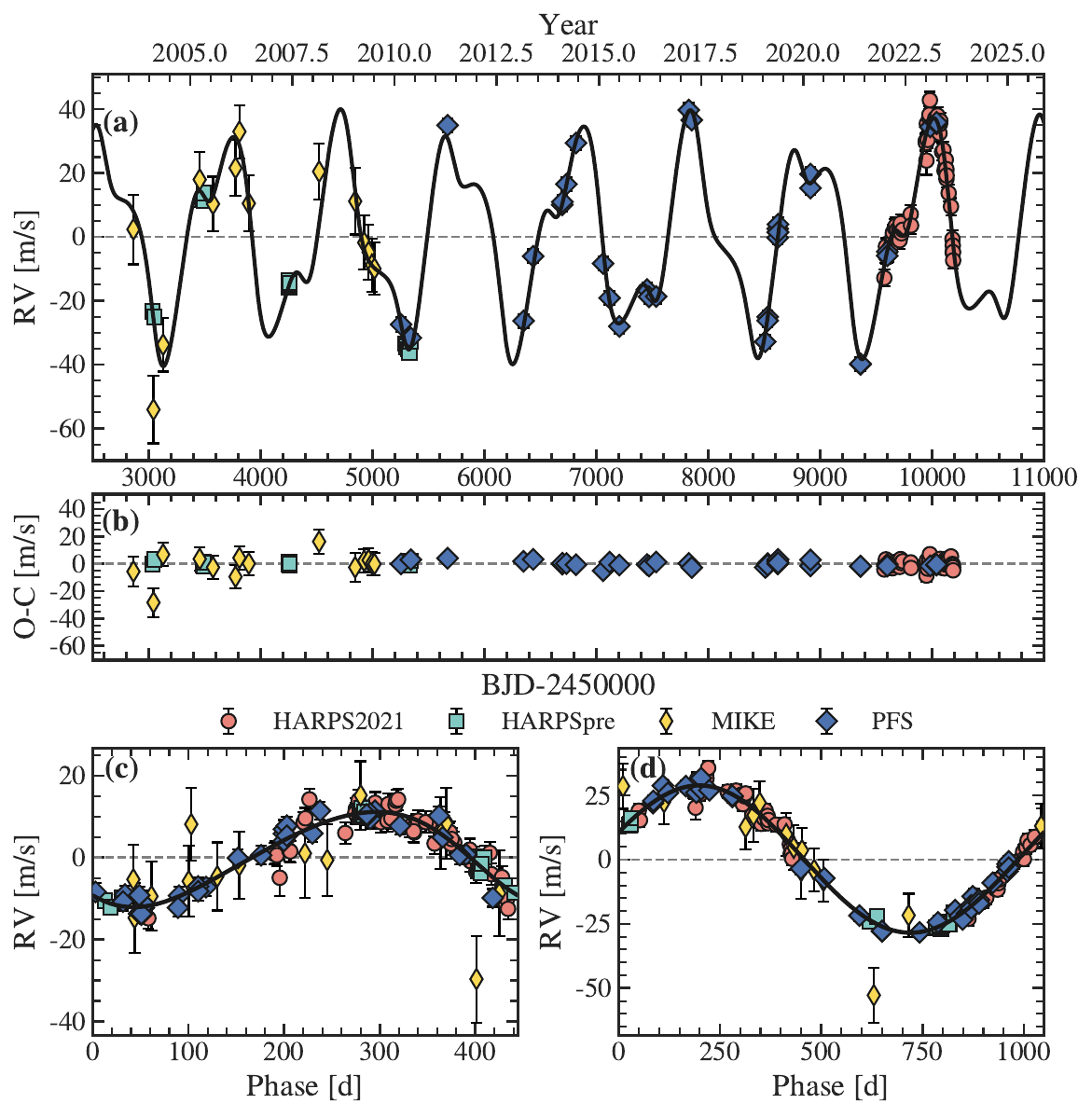}
    \caption{RV curve of HD 114386. The symbols are the same as Figure ~\ref{fig: HD48265RV}.}
    \label{fig: HD114386RV}
\end{figure}

\begin{figure}[htp]
    \centering
     \includegraphics[width=12cm]{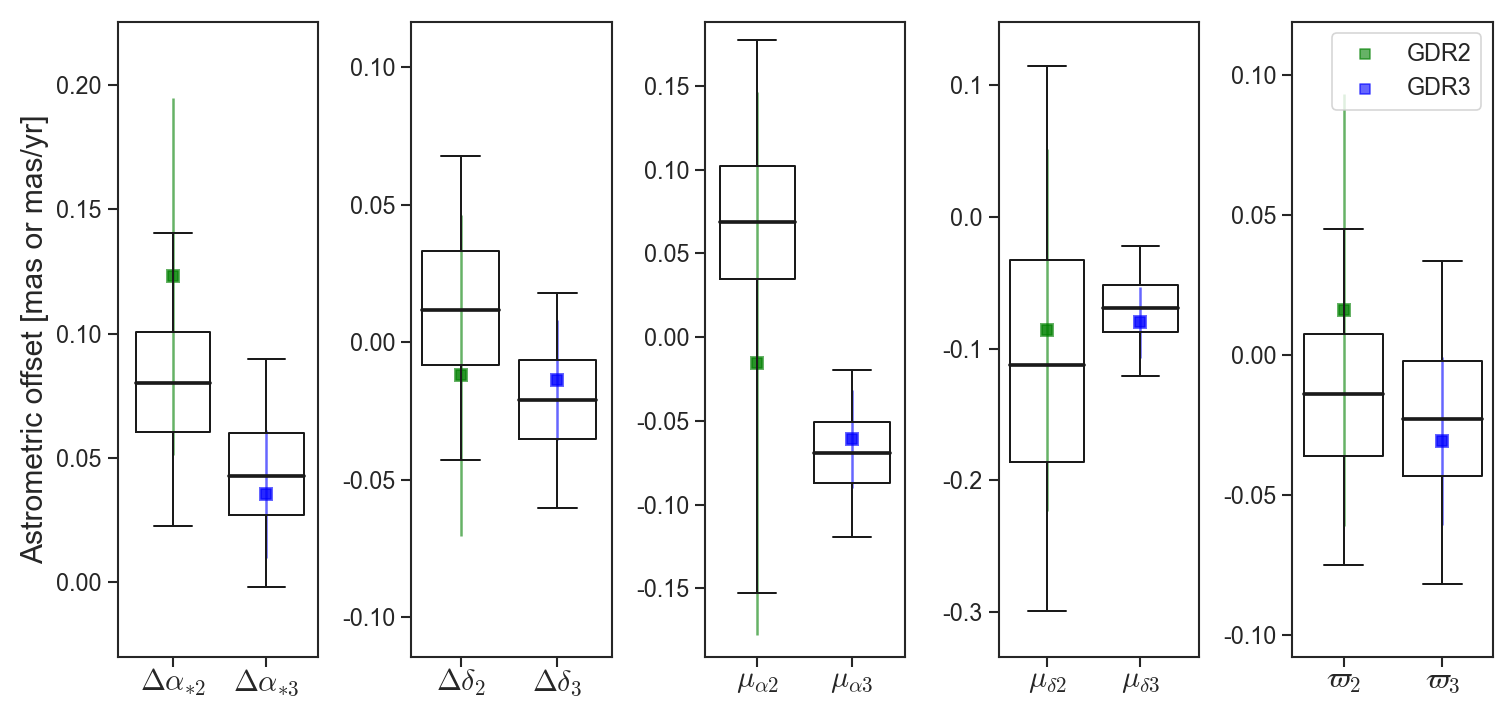}
        \caption{Boxplot for HD 114386. The symbols are the same as Figure ~\ref{fig: HD68475box}.}
        \label{fig: HD114386box}
\end{figure}

Specifically, as the naming in previous studies may have caused confusion, we refer to the inner planet as HD 114386 c and the outer planet as HD 114386 b, following the order of their discoveries. For HD 114386 c, due to its shorter period, astrometric sensitivity is limited, resulting in larger uncertainties in the orbital fit. We compare two models: Model A (ignoring the inner planet's astrometry for HD114386) and Model B (including it).  The computed Bayesian factor ($\ln \text{BF}_{\text{AB}} =(\text{BIC}_B - \text{BIC}_A)/2\approx 4$) indicates a positive preference for Model A over Model B, suggesting that ignoring the inner astrometry has a minor effect on the model's accuracy. Therefore, we conclude that the inner planet's astrometry in the HD114386 system can be safely ignored for this analysis. Based on RV data alone, the eccentricity is determined to be $ e_c=0.10\pm0.03$, with semi-major axis $a_c=1.05^{+0.04}_{-0.04}\, \mathrm{au}$. The minimum mass of this planet is constrained to be $M_\mathrm{c}\sin i=0.37^{+0.03}_{-0.03}\, M_\mathrm{Jup}$. The two signals in this system have surprisingly nearly circular orbits.
 The optimal RV and astrometric fits for HD 114386, based on our combined analysis, are shown in Fig.~\ref{fig: HD114386RV} and Fig.~\ref{fig: ast}, respectively. The latter also includes the planet's predicted motion on the sky. While Fig.~\ref{fig: HD114386box} 
 compares the five-parameter astrometry of the model prediction to GDR2 and GDR3 astrometry.

%%%%%%%%%%%%%%%%%HD 100508###########
\subsection{HD 100508}
\begin{figure}[htp]
    \centering
    \includegraphics[width=12cm]{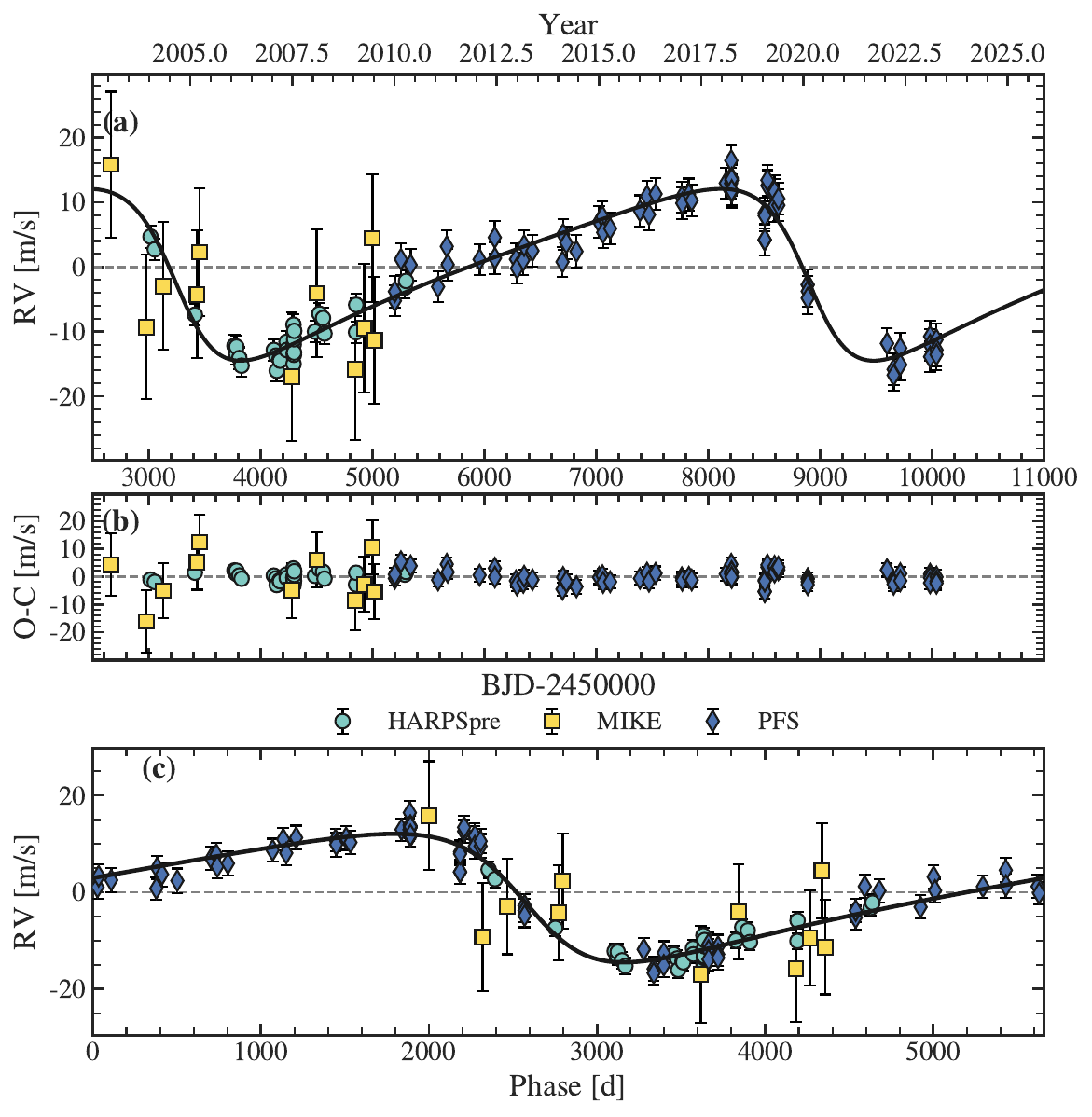}
        \caption{RV curve of HD 100508. The symbols are the same as Figure~\ref{fig: HD48265RV}.}
    \label{fig: HD100508RV}
\end{figure}

\begin{figure}[htp]
    \centering
 \includegraphics[width=12cm]{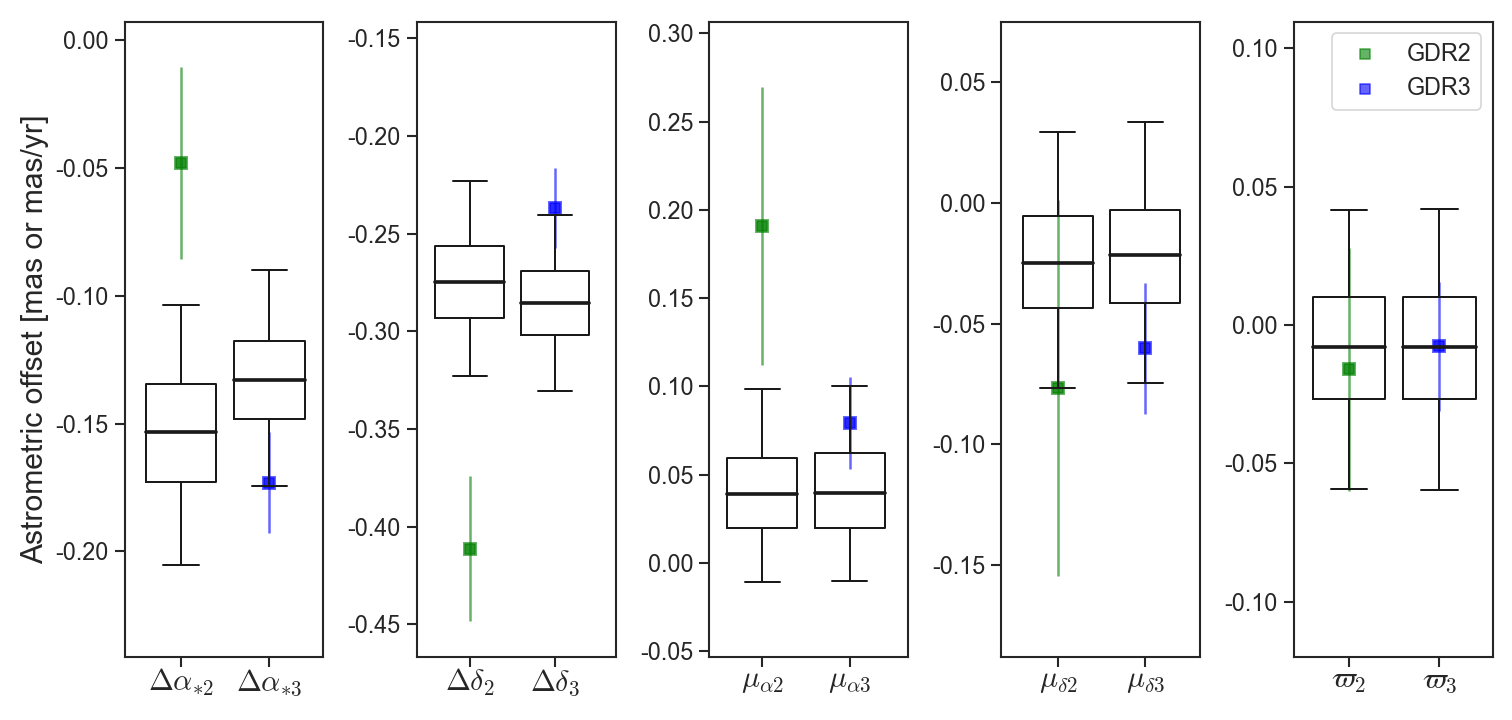}
        \caption{Boxplot of HD 100508. We note that several parameters exhibit $1\sim2 \sigma$ deviations between the Gaia measurements (GDR2/GDR3) and the model-predicted values. These offsets could stem from limitations in astrometric precision or the presence of a long-period companion that induces a subtle astrometric signal undetectable within the current Gaia baseline. Further analysis using more precise astrometry (e.g., the upcoming Gaia DR4) or multi-epoch modeling may help to resolve these discrepancies. Additional information is shown in Figure~\ref{fig: HD68475box}.}
    \label{fig: HD100508box}
\end{figure}

\begin{figure*}[htp]
    \centering
    \includegraphics[width=18cm]{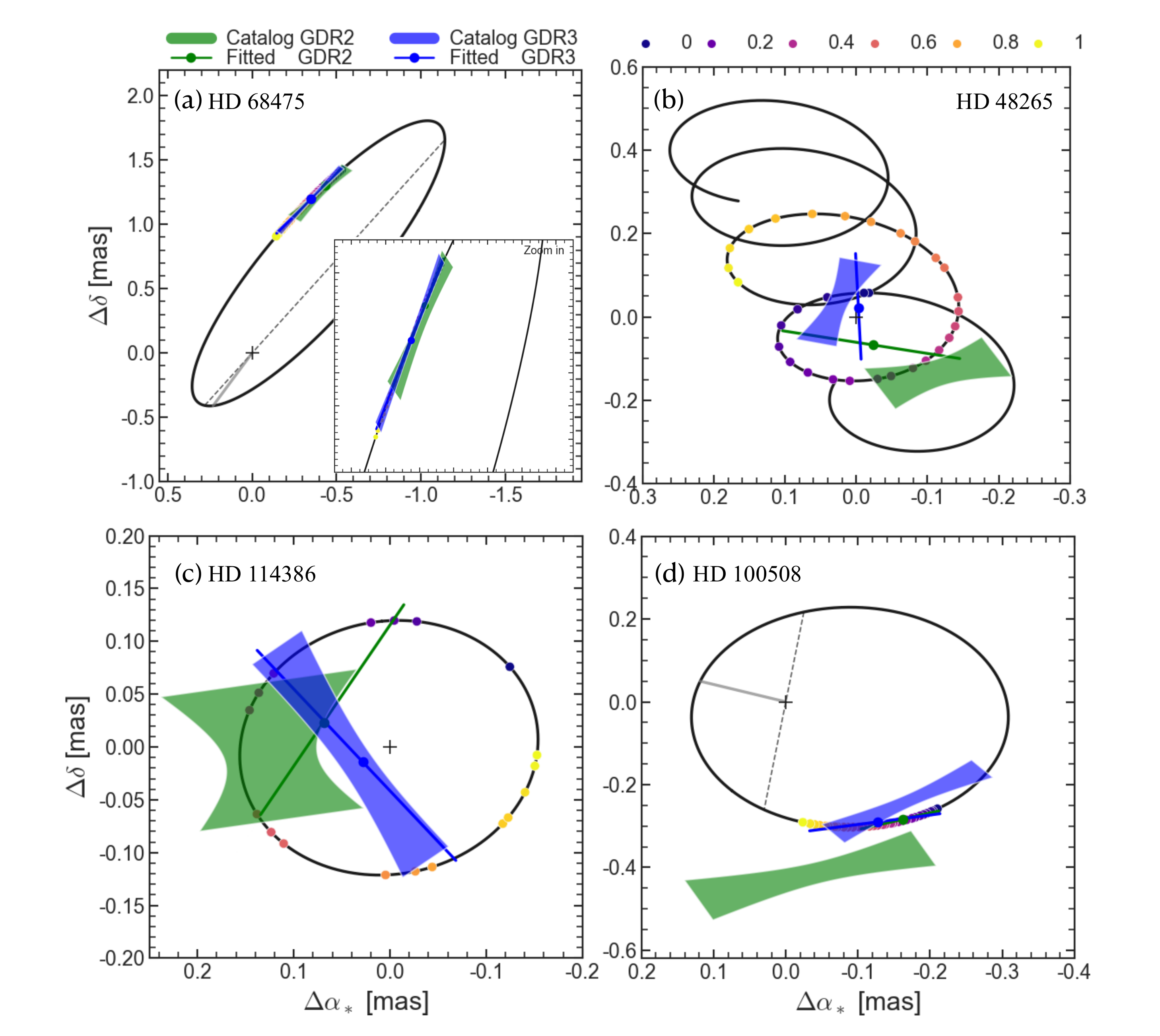}
        \caption{Astrometric fitting for four targets. (a) HD 68475: Best-fit astrometric orbit and a zoom-in on the Gaia observation region, showing the comparison between the model and catalog astrometry at the GDR2 and GDR3 reference epochs. The shaded regions represent uncertainties after TSB motion subtraction, and the dots and line slopes (blue and green) indicate the planet-induced offsets in position and proper motion. (b)–(d) Same as panel (a), showing the best-fit astrometric orbits for HD 48265, HD 114386, and HD 100508. The colored points represent binned model-predicted astrometric positions from the best-fit orbit, with color brightness increasing with time. The orbital uncertainties are further illustrated in Figure~\ref{fig: random}, where the time evolution of the relative astrometric offsets ($\Delta\alpha_*$ and $\Delta\delta$) is shown.
     \label{fig: ast}}
\end{figure*}

The host star of the HD 100508 planetary system is a K1 IV subgiant \citep{2006AJ....132..161G}. In this paper, we report the discovery of a new planet in this system. By fitting the RV data and combining it with astrometry data from GDR2 and GDR3, this planet is suggested to be a long-period cold Jupiter, with an orbital period of \( P_b = 15.56^{+0.12}_{-0.12} \, \text{yr} \), a semi-major axis of \( a_b = 6.11^{+0.25}_{-0.28} \, \text{au} \), and an eccentricity of \( e_b = 0.42^{+0.03}_{-0.03} \). The orbital inclination is found to be $i_b=62^{+19}_{-16}\, \mathrm{deg}$  or $i_b=124^{+14}_{-19}\ \mathrm{deg}$, with a mass of $M{_\text{b}}=1.20^{+0.30}_{-0.18} \, M\mathrm{_{Jup}}$. Other derived parameters include $a_b=6.11^{+0.25}_{-0.28}\ \mathrm{au}$. The longitude of the ascending node is \( \Omega_b = 174^{+34}_{-33} \, \mathrm{deg} \) for \( i_b > 90 \,\text{deg} \), and when \( i_b < 90\,
\text{deg}\), \( \Omega_b = 82^{+254}_{-58} \, \mathrm{deg} \) is obtained. %The two possible solutions for the orbital inclination arise because the current astrometry data are insufficient to resolve the system's center of mass. 
The two possible solutions for the orbital inclination arise because the current astrometry data are insufficient to break the degeneracy in the orbital orientation between prograde and retrograde configurations \citep{2024arXiv240308226F}.
Fig.~\ref{fig: HD100508RV} and Fig.~\ref{fig: ast} depict the optimal orbital solution for HD 100508 based on the MCMC posterior of RV+HG23. The former shows the best fit to RVs, while the latter shows the best fit to Hipparcos IAD and Gaia GOST data, and the predicted position of the planet.  As shown in Fig. \ref{fig: HD100508box}, the five-parameter astrometry derived from model prediction is compared against Gaia DR2 and DR3 measurements. Several parameters exhibit modest discrepancies, which could be attributed either to the inherent limitations in Gaia’s astrometric precision or to the influence of an undetected long-period companion inducing a small astrometric signal not yet resolved within the current Gaia time span.

\begin{figure*}[htp]
    \centering 
   \includegraphics[width=12cm]{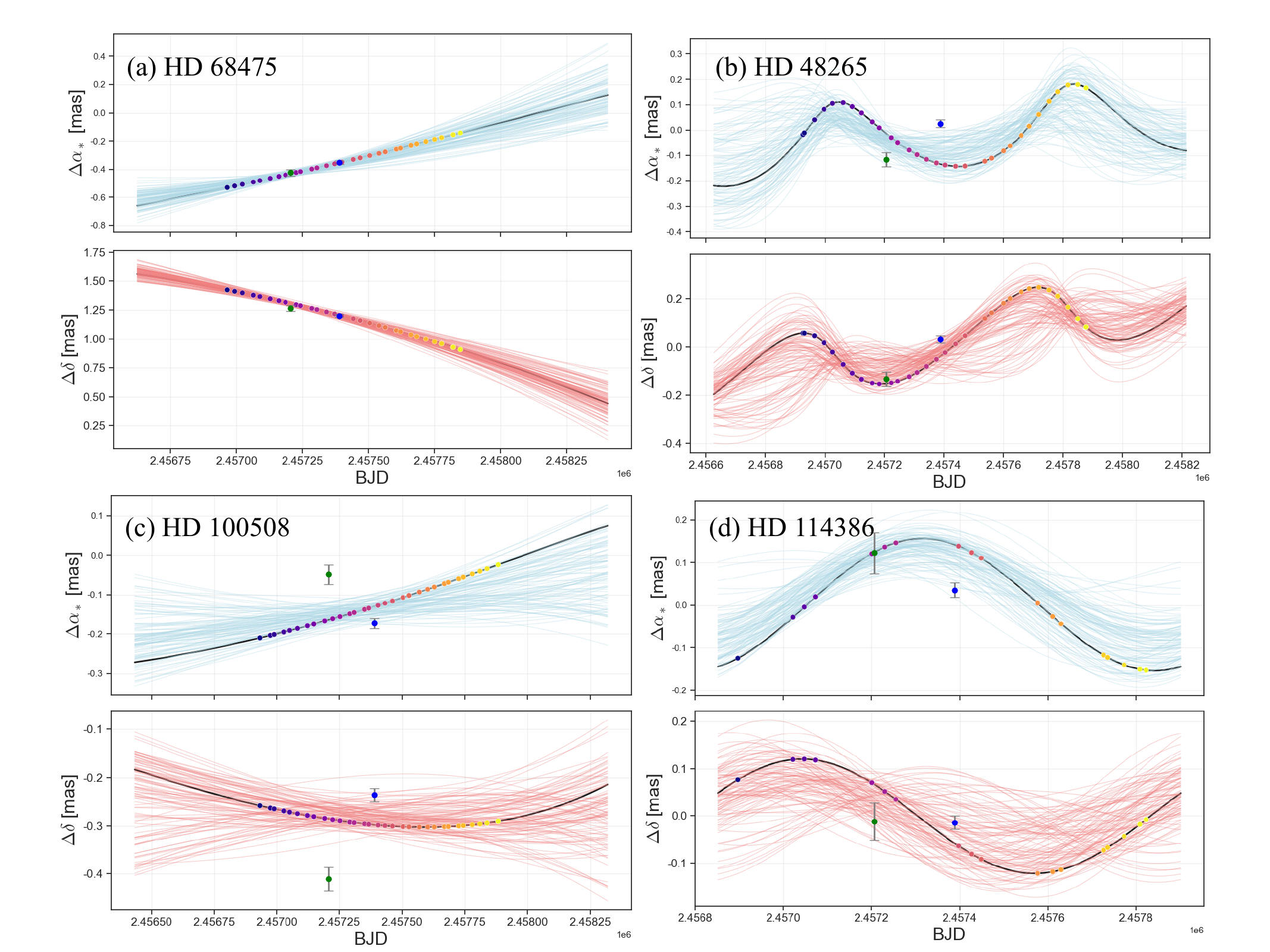}      
   \caption{Astrometric fitting of four targets as a function of the simulated observation time.
The black line shows the best-fit orbital model, while the thin colored lines represent randomly drawn samples from the posterior distributions, illustrating the uncertainty in the inferred orbits.
 } 
        \label{fig: random}
 \end{figure*}
 
\begin{figure*}[htp]
    \centering
    \includegraphics[width=16cm]{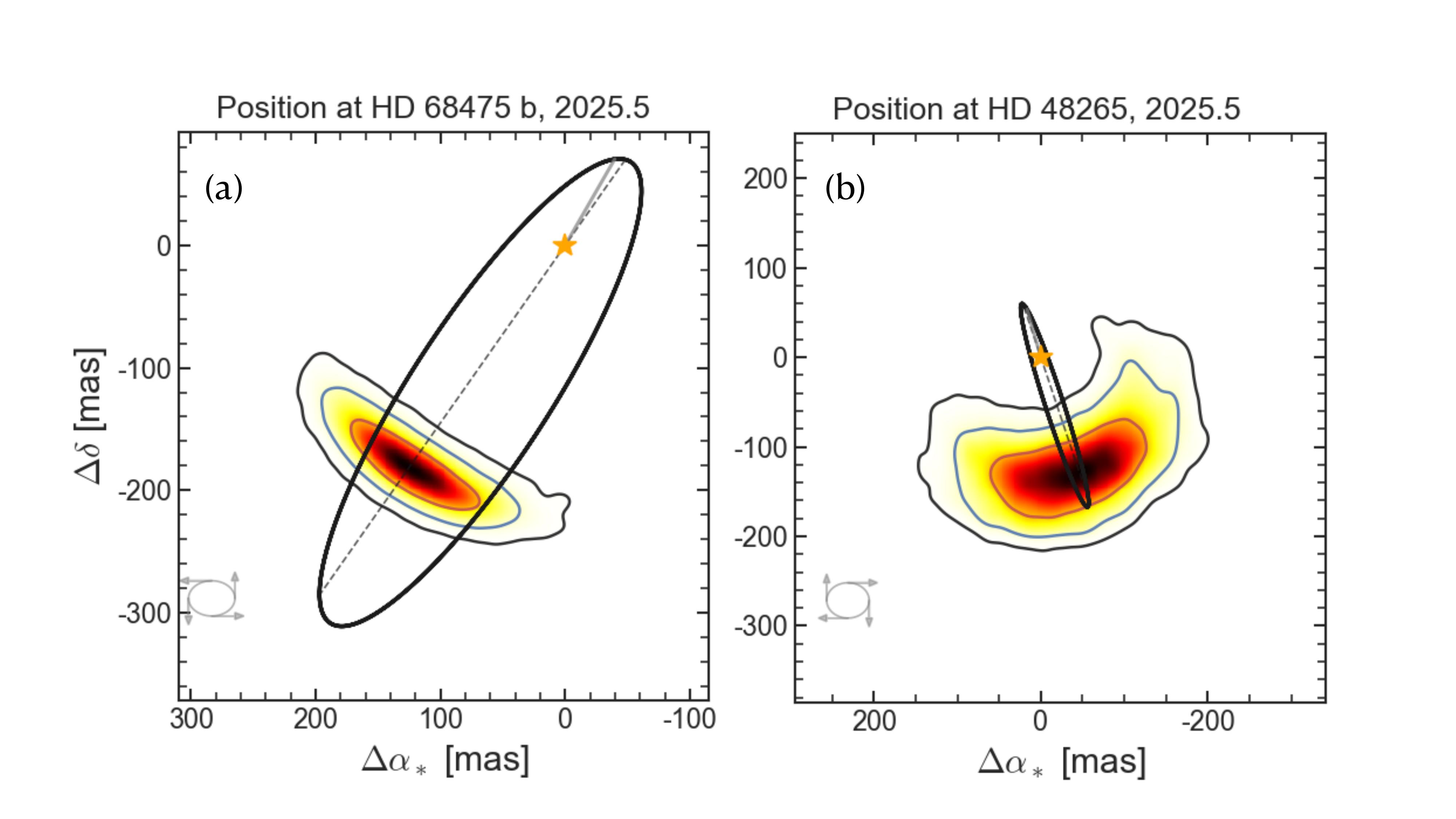}
        \caption{Predicted on-sky positions of HD 68475 b (a) and HD 48265 c (b) on July 2025, along with associated $1\sigma$, $2\sigma$, and $3\sigma$ uncertainties (contour lines). The curl at the lower left corner of each panel denotes the orientation of the orbital motion. For HD 68475 b, the predicted separation and position angle at epoch 2025 July 1 are $0.22 \pm 0.01$ arcsec and $146^\circ \pm 11^\circ$, respectively. For HD 48265 c, the predicted separation and position angle are $0.14 \pm 0.02$ arcsec and $298^\circ \pm 143^\circ$, respectively.
 }
     \label{fig: direct}
\end{figure*}

%***************parameters table**************
\begin{table}[htp]
\centering
\caption{Planetary parameters. The astrometric signals of HD 48265 b and HD 114386 c are not significant, while the orbital inclination of HD 100508 exhibits a bimodal distribution (see Figure~\ref{fig: 100508corner} in the appendix). This table lists the corresponding inclination (\(i\)) and longitude of ascending node (\(\Omega\)) for HD 100508 under the two possible solutions: \(i < 90^\circ\) and \(i > 90^\circ\). 
The last line shows the priors of our method. Log - $ \mathcal{U}(a, b)$ is the logarithmic uniform distribution between a and b, Cosi- $\mathcal{U} (a, b)$ is the cosine uniform distribution between a and b.
}
\hspace*{-3cm}
\resizebox{21cm}{!}{
\renewcommand{\arraystretch}{2}
\begin{tabular}{lllllllllllllll}
\toprule
\hline
Name & $M_\text{p}$& $i $  & $a$  & $e$  & $P$  & $\Omega$ & $\omega$  & $T_\text{p}$  & $M_0$& $M_\text{p}\sin i$ \\
 & ($M_{\text{Jup}}$)& ($^\circ$)  & (au) &   & (day)  &($^\circ$) & ($^\circ$) & day & ($^\circ$)&($M_{\rm Jup}$)&  \\
 
\hline
HD 48265 b  &${7.4}_{-3.9}^{+3.7}$ &${11.7}_{-3.4}^{+11.0}\,(i\textless\, 90^{\circ})$ &   ${1.87}_{-0.04}^{+0.02}$  & ${0.35}_{-0.02}^{+0.02}$ & ${789.6}_{-1.1}^{+1.1}$    & ${40}_{-15}^{+19}\,(i\textless\, 90^{\circ})$   & ${34.5}_{-3.0}^{+3.1}$  & ${2452275.6}_{-9.1}^{+8.9}$ &${294.2}_{-3.7}^{+3.8}$&$1.67_{-0.06}^{+0.06}$&  \\

&& ${160.1}_{-31}^{+8.3}\,(i \textgreater 90^{\circ})$&&&&${80}_{-27}^{+26}\,(i \textgreater 90^{\circ})$&&&\\

HD 48265 c  & ${4.45}_{-0.37}^{+0.75}$ & ${89}_{-29}^{+29}$ &${10.4}_{-1.0}^{+1.6}$     & ${0.41}_{-0.06}^{+0.08}$ & ${10418}_{-1412}^{+2451}$  & ${196}_{-26}^{+27}$     & ${24.0}_{-3.9}^{+4.1}$  & ${2446077}_{-2568}^{+1317}$ & ${240}_{-18}^{+24}$&$4.09_{-0.20}^{+0.24}$ \\

HD 68475 b  & ${5.16}_{-0.47}^{+0.53}$ & ${87}_{-19}^{+21}$ &     ${7.27}_{-0.36}^{+0.39}$    & ${0.62}_{-0.02}^{+0.02}$ & ${7832}_{-323}^{+463}$     & ${327.2}_{-8.4}^{+8.4}$     & ${195.6}_{-1.7}^{+1.8}$ & ${2451564}_{-466}^{+325}$   & ${65}_{-13}^{+17}$&$4.90_{-0.40}^{+0.40}$ \\

HD 100508 b & ${1.20}_{-0.18}^{+0.30}$ & ${62}_{-16}^{+19} \,(i\textless\, 90^{\circ})$ & ${6.11}_{-0.28}^{+0.25}$    & ${0.420}_{-0.02}^{+0.02}$ & ${5681}_{-42}^{+42}$       & ${82}_{-58}^{+254}\,(i \textless90^{\circ})$& ${99.1}_{-4.2}^{+4.1}$  & ${2447556}_{-84}^{+87}$     & ${323.4}_{-3.7}^{+3.6}$&$1.02_{-0.09}^{+0.09}$\\

&& ${124}_{-19}^{+14}\,(i \textgreater 90^{\circ})$&&&&${174}_{-33}^{+34}\,(i \textgreater 90^{\circ})$&&&\\

HD 114386 b & ${1.46}_{-0.22}^{+0.37}$ & ${57}_{-15}^{+22}$ &                   ${1.86}_{-0.08}^{+0.07}$ &  ${0.02}_{-0.01}^{+0.01}$ & ${1049.4}_{-1.2}^{+1.5}$   &   ${278}_{-24}^{+19}$                                                                                         &   ${58}_{-37}^{+270}$   &  ${2452697}_{-139}^{+109}$    &   ${57}_{-38}^{+48}$&$1.22_{-0.10}^{+0.10}$  \\
HD 114386 c & -- &  --  &                      ${1.05}_{-0.04}^{+0.04}$ &   ${0.10}_{-0.03}^{+0.03}$ & ${444.00}_{-0.88}^{+0.93}$ &   --   &  ${123}_{-16}^{+18}$   &  ${2452626}_{-18}^{+21}$    &  ${192}_{-17}^{+14}$& $0.37_{-0.03}^{+0.03}$ \\
Prior &-- & Cos i - $\mathcal{U}(-1,1)$ &-- &$\mathcal{U}(0,1)$ &Log - $\mathcal{U}(-1,16)$ &$\mathcal{U}(0,2\pi)$ &$\mathcal{U}(0,2\pi)$ &-- &$\mathcal{U}(0,2\pi)$ &--\\
\hline
\bottomrule
\end{tabular}}
\end{table}

Table~\ref{tab:rv_summary} lists a representative sample of the radial velocity measurements used in this study. The full RV dataset we used, including all available epochs for each instrument, is provided in machine-readable format in a separate file.

\begin{table*}[htbp]
\centering
\caption{Excerpt of Radial Velocity Measurements}
\label{tab:rv_summary}
\begin{tabular}{lllll}
\toprule
\hline
Signal name   & BJD         & RV           & eRV         & Instrument \\
& & $\rm{m/s}$&$\rm{m/s}$&\\
\hline
HD 48265  & 2452920.863 & 7.58         & 6.95        & MIKE       \\
         & 2452977.777 & 9.52         & 7.52        & MIKE       \\
         & 2453431.607 & -41.84       & 2.74        & MIKE       \\
         & $\vdots$ & $\vdots$       & $\vdots$     & $\vdots$      \\
         
\hline
\bottomrule
\end{tabular}
\end{table*}
\section{Discussion}\label{sec:discussion}
\subsection{Possible target for direct imaging}

HD 68475 b is a cold Jupiter candidate orbiting a K2V-type host star,  $T_{\text{eff}}$ = $5022\pm113$ K.  
Using isochrones fitting based on stellar parameters (e.g., \( T_{\mathrm{eff}} \), \(\log g\), and [Fe/H]) \citep{2005A&A...436..127J}, we estimate a stellar age of approximately $5 \pm1$ Gyr. The effective temperature of the companion is then estimated to be \(240 \pm 40\,\mathrm{K}\) by matching this age with the ATMO 2020 atmospheric models \citep{2020A&A...637A..38P}, which correspond to a planetary mass of \(\sim 0.005\,M_{\odot}\).

Based on its expected flux, HD~68475~b exhibits a planet-to-star contrast of approximately $1.79 \times 10^{-5}$ at the F1140C band, $4.4 \times 10^{-4}$ at the F1550C band, and $1.92 \times 10^{-3}$ at the F2300C band of JWST. While the planet's contrast at the F1550C and F2300C bands is brighter than JWST/MIRI's nominal detection limits, the inner working angles (IWAs) at these wavelengths are too large to resolve the planet’s projected separation ($\sim 0.22''$, see Fig.~\ref{fig: direct}).
The F1140C band provides a smaller IWA ($\sim 0.36''$), yet the contrast at this wavelength falls below the instrument’s detection threshold \citep{2022A&A...667A.165B}. As a result, HD~68475~b remains inaccessible to JWST for direct imaging. 
However, future observations in the $N$ band with facilities offering both high contrast sensitivity and finer angular resolution may offer a promising path forward.

The Extremely Large Telescope (ELT), equipped with the Mid-infrared ELT Imager and Spectrograph (METIS), is expected to significantly advance the direct imaging capabilities for exoplanets in the thermal infrared range ($8$–$13~\mu$m) \citep{2021Msngr.182...22B}. METIS utilizes advanced coronagraphic techniques, including a vortex phase mask, and is expected to achieve an inner working angle (IWA) of approximately $2.5\,\lambda/D$ in the N2 band (central wavelength $11.2\mu $m), which would correspond to about $148\ \mathrm{mas}$ for the ELT's 39-meter aperture.

In this wavelength range, the expected planet–star contrast for HD~68475~b is comparable to that in the JWST/MIRI F1140C band. This IWA is well below the projected angular separation of HD 68475 b ($\sim217~\mathrm{mas}$), indicating that METIS could easily resolve and potentially image such a mature cold giant in thermal emission.

\subsection{Dynamical Properties and Orbital Architectures of the Systems}
\subsubsection{HD 48265}
The HD 48265 system consists of two confirmed giant planets, with orbital periods of $789\pm1.1$ d and $10418^{+2451}_{-1412}$d, respectively. Orbital fitting yields a significant mutual inclination, suggesting that the planets are on strongly misaligned orbits. 
%While the derived orbital elements are subject to large uncertainties — especially for the inner companion, given the current limitations of Gaia DR3 astrometry — the available data allow for tentative inferences about the system's dynamical history. 
While the astrometric data generally provide useful constraints, for some planets — particularly those with short orbital periods or limited data — the inclination is not well constrained, and the data do not provide the level of significance expected for strong constraints. Despite these uncertainties, the available data still allow for tentative inferences about the system's dynamical history.

%The contribution of inner planets is negligible because of their small semi-major axis. However, these parameters remain of interest.
The mutual inclination \(\Phi\) between the two orbital planes is estimated using the following expression:
\begin{equation}
\cos \Phi = \cos i_b \cos i_c + \sin i_b \sin i_c \cos(\Omega_b - \Omega_c)
\end{equation}
\begin{figure*}[htp]
   \centering 
   \includegraphics[width=15cm]{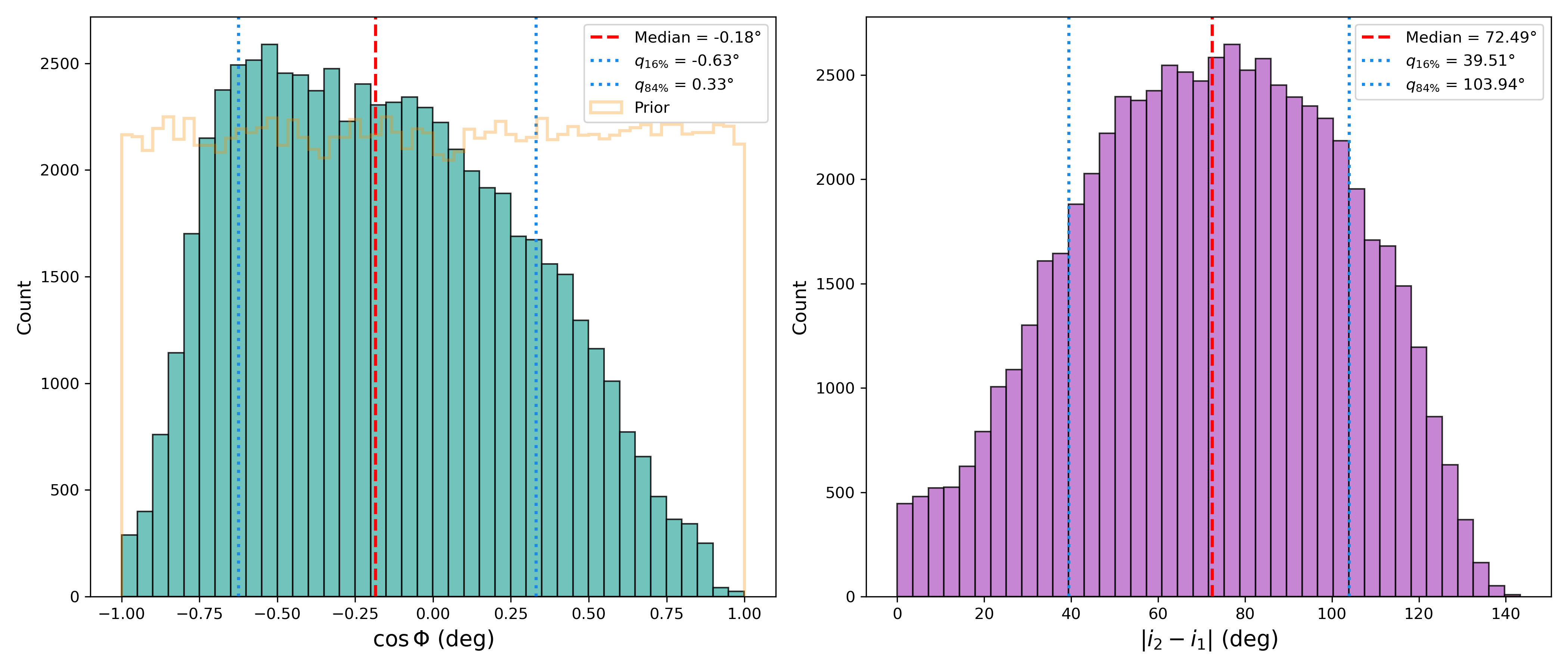}      
   \caption{Left panel: Posterior distribution of $\cos \Phi$ for the mutual inclination $\Phi$ between HD 48265 b and c. The orange curve represents the prior distribution assumed for the mutual inclination, which is uniformly sampled and thus influenced by the alignment of the system ($\Omega$). The red dashed line indicates the median value ($-0.18$), while the blue dotted lines mark the 16th ($-0.63$) and 84th percentiles ($0.33$). Converting $\cos \Phi$ to $\Phi$ yields an inclination of approximately $100.4^\circ$.
   Right panel: The distribution of the absolute value of the minimum mutual inclination ($|i_2 - i_1|$). Since no prior on $\Omega$ is considered here, the results reflect the lowest mutual inclination angle for the system. The median value is $72^\circ$, with the 16th and 84th percentiles at $40^\circ$ and $104^\circ$, respectively. Zero degrees is excluded at more than $2\sigma$ significance, indicating that the system is significantly mutually inclined.}
       \label{fig: mutual_i}
\end{figure*}
As shown in Figure~\ref{fig: mutual_i}, the posterior distribution of the mutual inclination (\(\Phi\)) peaks near the median value of \(100.4^\circ\). This large inclination suggests that the two planets may have undergone significant dynamical excitation if they were initially born in a protoplanetary disk.

\begin{figure*}[htp]
   \centering 
   \includegraphics[width=12cm]{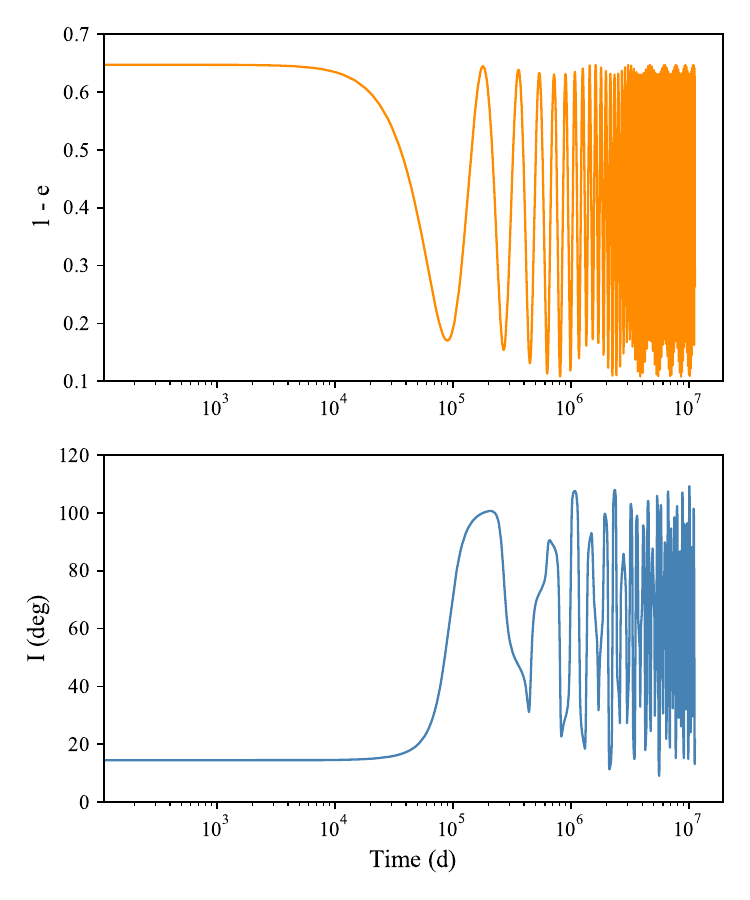}      
   \caption{Evolution of the planetary orbit as a function of  time. The top panel represent the evolution of the inner planet eccentricity and the bottom panel shows the evolution of the inner planet orbital inclination.}
       \label{fig: kozai}
\end{figure*}

By analogy with the HAT-P-11 system \citep{2024ESS.....561604L}, a possible formation pathway for HD 48265 is speculated to involve an early episode of planet-planet scattering.  Planet-planet scattering refers to close gravitational encounters between planets that can lead to ejections or changes in eccentricity and inclination. In this scenario, the original system may have contained additional planets, several of which were dynamically ejected following gravitational interactions, and two planets were left \citep{2008ApJ...686..621F}. The remaining planets are on misaligned and eccentric orbits, with two planets undergoing von-Zeipel-Kozai-Lidov (KL) oscillations, which drive their orbital evolution. These KL oscillations, arising from a high mutual inclination system ($40^\circ \textless \Phi \textless 140^\circ$), cyclically exchange the eccentricity and inclination of the inner planet, and the observed high eccentricity likely corresponds to a specific phase within the KL cycle. By simulating the HD 48265 system with REBOUND using the parameters from our previous work, we adopted the measured eccentricity, among other parameters, as the initial condition. The system exhibits KL-type periodic behavior, as demonstrated by the orbital evolution of inner planet shown in Figure ~\ref{fig: kozai}, plotted as a function of logarithmic time (in days).
It is important to note that the observed eccentricity of this system represents only a snapshot at a specific phase. If the high eccentricity persists over time, additional effects such as tidal circularization, mass loss, dynamical instability, or scattering by a perturber may arise.

HD 48265 presents a promising case for investigating how planet-planet scattering and von-Zeipel-Kozai-Lidov migration may influence the orbital architecture of planetary systems. The current orbital solutions, especially for the inner planet, represent preliminary estimates that will benefit from further refinement with future Gaia data releases and continued observational efforts. These advancements will help clarify the mutual inclination and the dynamical history of the system.

\subsubsection{HD 114386}
The orbital eccentricities of planets b and c in the HD 114386 system are approximately $e_b = 0.02^{+0.01}_{-0.01}$,$P_b=1049.4^{+1.5}_{-1.2}$ and $e_c = 0.10^{+0.04}_{-0.04}$, $P_c=444.8^{+0.4}_{-0.4}$, respectively. While HD 114386 b exhibits a nearly circular orbit, HD 114386 c shows a modestly higher eccentricity. The low eccentricity of planet b is likely due to strong eccentricity damping via dynamical friction from the gas disk during the protoplanetary phase \citep{2000MNRAS.315..823P, 2003ApJ...585.1024G}. Its high mass ($\sim 1.46 \, M_{\text{Jup}}$) suggests that it likely underwent Type II migration, in which it opened a gap in the disk and migrated slowly, further contributing to eccentricity suppression \citep{1986ApJ...309..846L, 2012ARA&A..50..211K}. In contrast, HD 114386 c ($M_{\text{c}}\sin i = 0.37^{+0.03}_{-0.03} \, M_{\text{Jup}}$) lies near the transition region between Type I and Type II migration regimes. Its slightly elevated eccentricity could reflect a more complex migration history, potentially involving planet-disk interactions that were less efficient in damping, or post-disk dynamical interactions with planet b. Its specific migration mechanism may depend on local disk properties such as surface density and disk temperature \citep{2004ApJ...602..388T}.

 %%%%%%%%HD 100508 和HD 68475 %%%%%%%%
 
\begin{figure*}[htp]
   \centering 
   \includegraphics[width=15cm]{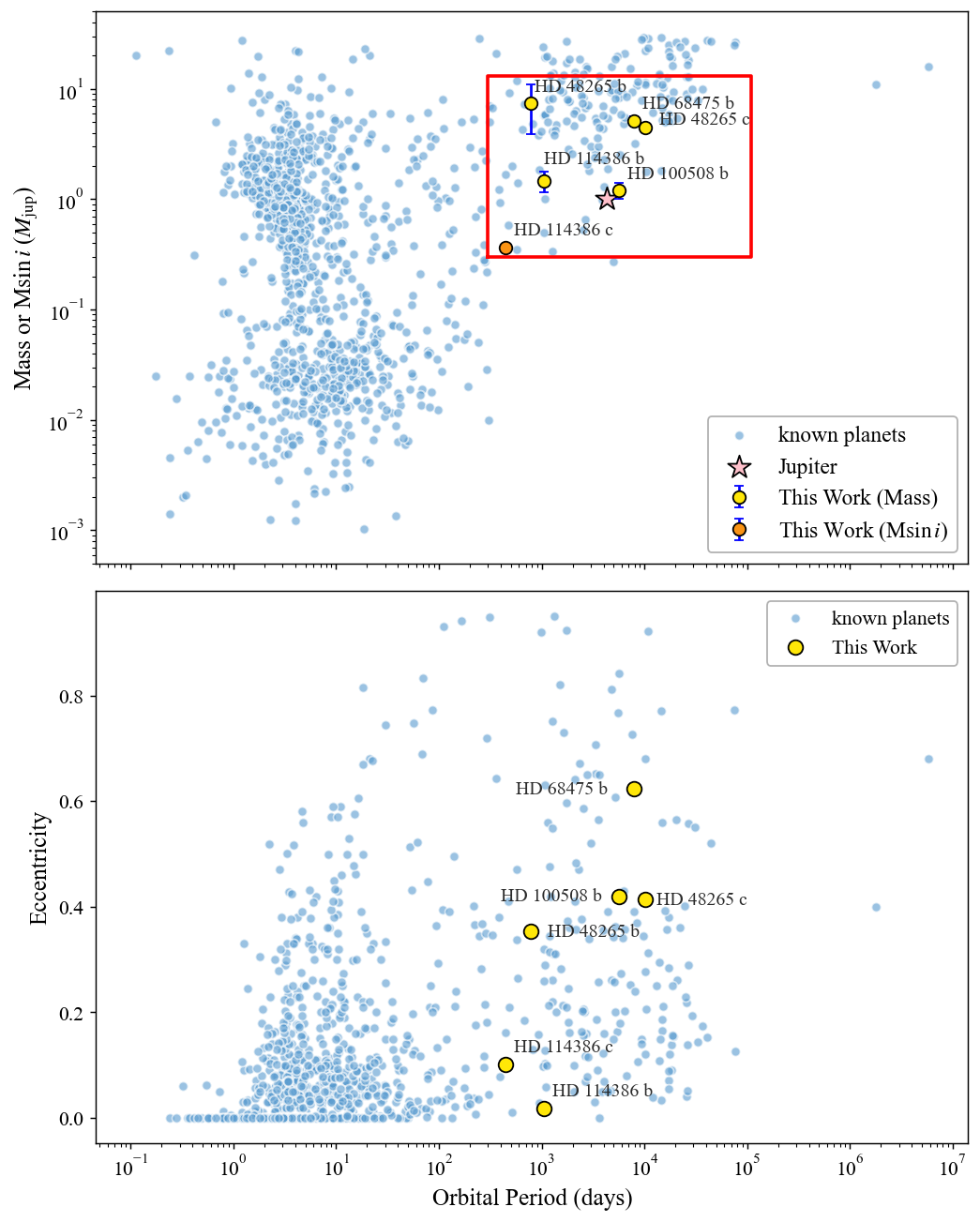}      \caption{Distribution of period--mass (P--M, top) and period--eccentricity (P--e, bottom).  Blue circles represent a selected subset of known exoplanets from the NASA Exoplanet Archive \citep{akeson2013nasa} with measured orbital inclinations, allowing determination of true masses. Yellow stars indicate the planets studied in this work. Due to limited observational precision, the true masses of  HD 114386 c are not well constrained; It is therefore marked with orange stars in the period--mass diagram. The red rectangle marks the typical region occupied by cold Jupiters.} 
       \label{fig: P_M plot}
\end{figure*}

Figure~\ref{fig: P_M plot} shows the distribution of the planets reported in this paper on the period--mass diagram. Among them, HD~100508~b and HD~68475~b stand out due to their relatively long orbital periods. 
HD~100508 hosts a jupiter analog with a mass of $1.2\, M_{\mathrm{Jup}}$ and an orbital period of 5681~d, which lies at the long-period end of currently known planets within the $0.5\sim 2\, M_{\mathrm{Jup}}$ mass range.
 HD~68475~b, with a more massive planet ($5.16\, M_{\mathrm{Jup}}$) and a period of 7832~d, exhibits an even higher eccentricity of 0.6. While no additional companions have been detected in either system, the eccentricity of HD~68475 may point to the possibility of past dynamical interactions.

\section{Conclusion}
\label{conclu}

This work combines radial velocity and astrometry methods, using high-precision RV data from the PFS project and other publicly available archival RV datasets, along with astrometry data from Gaia and Hipparcos. By employing the \texttt{ptemcee} tool to iteratively solve the orbital equations of long-period cold Jupiters and astrometric parameters, highly reliable orbital parameters, as well as the true masses and orbital inclinations of several sources, are ultimately obtained. 

We report the discovery of four planetary companions orbiting stars that span spectral types from G5 to K3, encompassing both main-sequence and subgiant stages. HD 100508 b and HD 68475 b represent the first planetary signals detected in their respective stellar systems, marking them as newly identified single-planet systems. For HD 100508 b, the orbital period is $5681^{+42}_{-42}$ d, and the mass is $1.20^{+0.30}_{-0.18}\,M_\text{Jup}$. HD 68475 b has an orbital period of $7831^{+463}_{-323}$ d, with a mass of $5.16^{+0.53}_{-0.47} \,M_\text{Jup}$. In contrast, the systems of HD 48265 and HD 114386 already host previously confirmed planets. In these systems, evidence for additional, yet unconfirmed planetary signals is found, which are designated as HD 48265 c and HD 114386 c. HD 48265 c is an outer planet with a longer orbital period of $10418^{+2400}_{-1300}$ d, while HD 114386 c is an inner planet with a period of $444.8^{+0.4}_{-0.4}$ d. The mass of HD 48265 c is $3.71^{+0.68}_{-0.43} \,M_\text{Jup} $ and HD 114386 c has a minimum mass of $M_\text{c} \sin i=0.37^{+0.03}_{-0.03}\,M_\text{Jup}$.   
  We also update the orbital parameters and true masses or minimum mass of previously confirmed stars like HD 48265 b and HD 114386 b.

This study expands the database of cold Jupiters and enhances our understanding of planetary diversity and the complexity of system architectures. The high mutual inclination and elevated eccentricity of HD 48265 could suggest potential histories involving planet-planet scattering or von-Zeipel–Kozai–Lidov cycles, while the nearly circular orbit of HD 114386 b may be attributed to migration processes and dynamical friction from the gas disk. The high eccentricities observed in HD 68475 and HD 100508 might reflect dynamical perturbations from undetected companions, and HD 68475 b’s wide orbital separation could make it a candidate for future direct imaging efforts. These systems collectively provide important references for investigating the formation mechanisms, internal structures, and long-term orbital evolution of cold giant planets.

In 2026, Gaia DR4  will provide an unprecedented volume of high-precision astrometric and time-series data on exoplanets, which will significantly improve the determination of orbital parameters and support long-term dynamical studies. In parallel, direct detection methods like high-contrast imaging will further expand our knowledge of planetary properties and system configurations. As astronomical observation techniques continue to advance, particularly in the time-domain and high-precision astrometry, our understanding of the formation and long-term dynamical evolution of cold Jupiters is expected to deepen.

\section*{Acknowledgments}
We thank Xiumin Huang and Kangrou Guo for their valuable suggestions on the dynamical analysis, and Yifan Xuan, Shuyue Zheng, Yicheng Rui for their helpful comments on the manuscript. This work is supported by the National Key R\&D Program of China, No. 2024YFA1611801 and No. 2024YFC2207700, by the National Natural Science Foundation of China (NSFC) under Grant No. 12473066, by the Shanghai Jiao Tong University 2030 Initiative, and by the China-Chile Joint Research Fund (CCJRF No. 2205). CCJRF is provided by the Chinese Academy of Sciences South America Center for Astronomy (CASSACA) and established by the National Astronomical Observatories, Chinese Academy of Sciences (NAOC), and Chilean Astronomy Society (SOCHIAS) to support China-Chile collaborations in astronomy. The authors acknowledge the years of technical support from LCO staff in the successful operation of PFS, enabling the collection of the data presented in this paper. 
The computations in this paper were run on the $\pi$ 2.0 (or the Siyuan-1) cluster supported by the Center for High Performance Computing at Shanghai Jiao Tong University. This work is partly based on observations collected at the European Organisation for Astronomical Research in the Southern Hemisphere under ESO programs: 108.22KV, 60.A-9036, 072.C-0488, 079.C-0681, 087.D-0511, 091.C-0844, 094.C-0894, 192.C-0852, 196.C-1006, 105.20MP.001, 108.2271.002 and 108.2271.003. Part of this research was carried out at the Jet Propulsion Laboratory, California Institute of Technology, under a contract with the National Aeronautics and Space Administration (80NM0018D0004).
\software{
We also acknowledge the use of several open-source software tools that facilitated the analysis and visualization in this work. Astropy \citep{2013A&A...558A..33A}, NumPy \citep{harris2020array}, SciPy \citep{virtanen2020scipy}, pandas \citep{2020zndo...3898987R}, matplotlib \citep{hunter2007matplotlib}, seaborn \citep{waskom2021seaborn}, corner \citep{foreman2016corner}, and ptemcee \citep{2016MNRAS.455.1919V}.
In particular, we thank the developers and maintainers of \texttt{Anaconda} for providing a comprehensive and flexible package management system, \texttt{Jupyter Notebook} for its user-friendly and interactive computing environment, and \texttt{Spyder} for its integrated development environment optimized for scientific programming in Python.}

 %This work is based in part on data acquired at the Anglo-Australian Telescope. We acknowledge the traditional custodians of the land on which the AAT stands, the Gamilaraay people, and pay our respects to elders past and present. 

\bibliographystyle{aasjournal}
\bibliography{ref}

\appendix
\begin{figure}[htp]
    \centering
    \includegraphics[width=18cm]{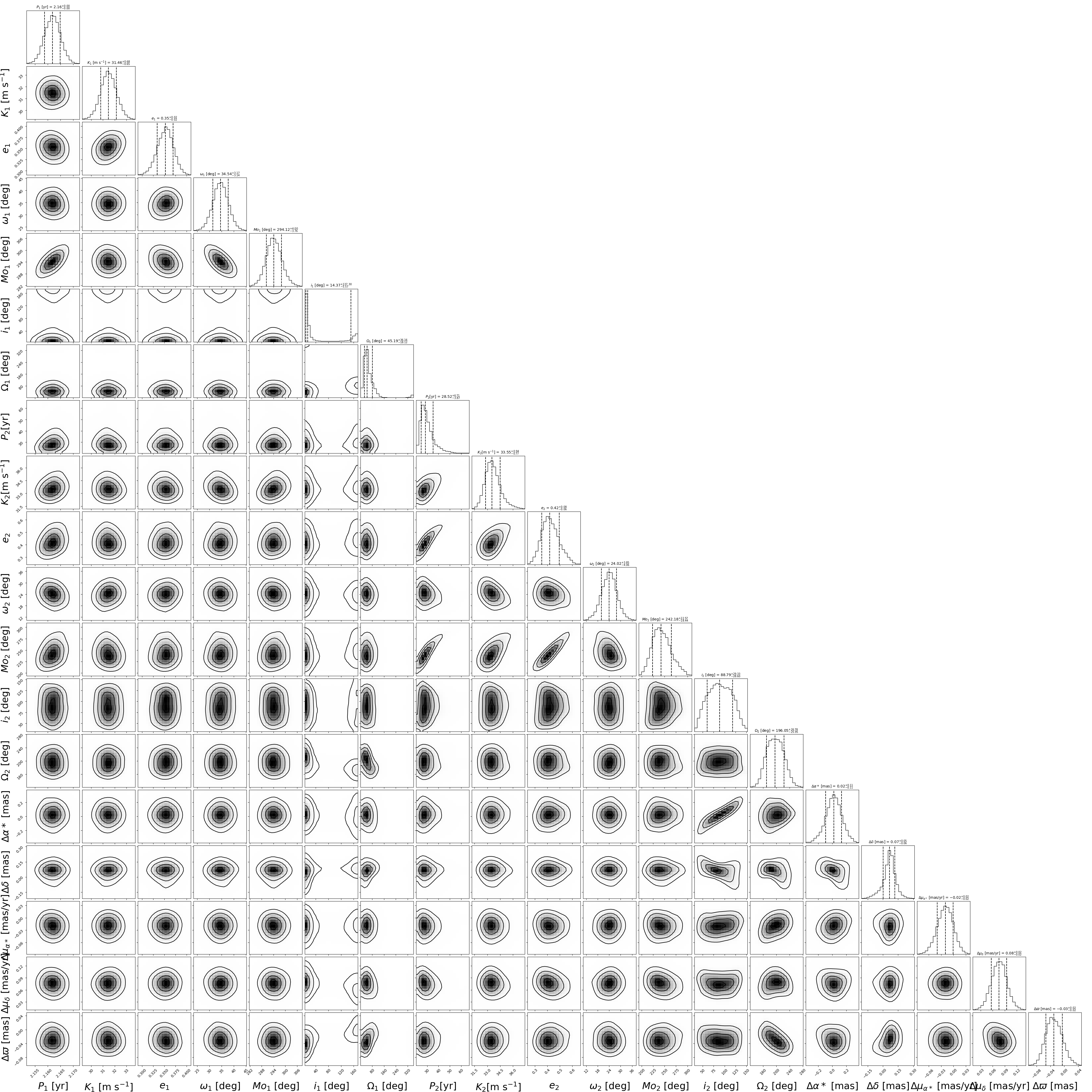}
    \caption{Corner plot showing the posterior distributions of HD 48265 b and c from the joint RV and astrometry fit. The parameters include orbital period ($P$), RV semi-amplitude ($K$), eccentricity ($e$), inclination ($i$), argument of periastron ($\omega$), longitude of ascending node ($\Omega$), astrometric offsets ($\Delta\alpha$, $\Delta\delta$), proper motion offsets ($\Delta\mu_\alpha$, $\Delta\mu_\delta$), and parallax offset ($\Delta\pi$). Diagonal panels show the marginalized 1D posteriors, with vertical dashed lines indicating the median and 1$\sigma$ credible intervals. }
    \label{fig: 48265corner}
\end{figure}

\begin{figure}[htp]
    \centering
    \includegraphics[width=18cm]{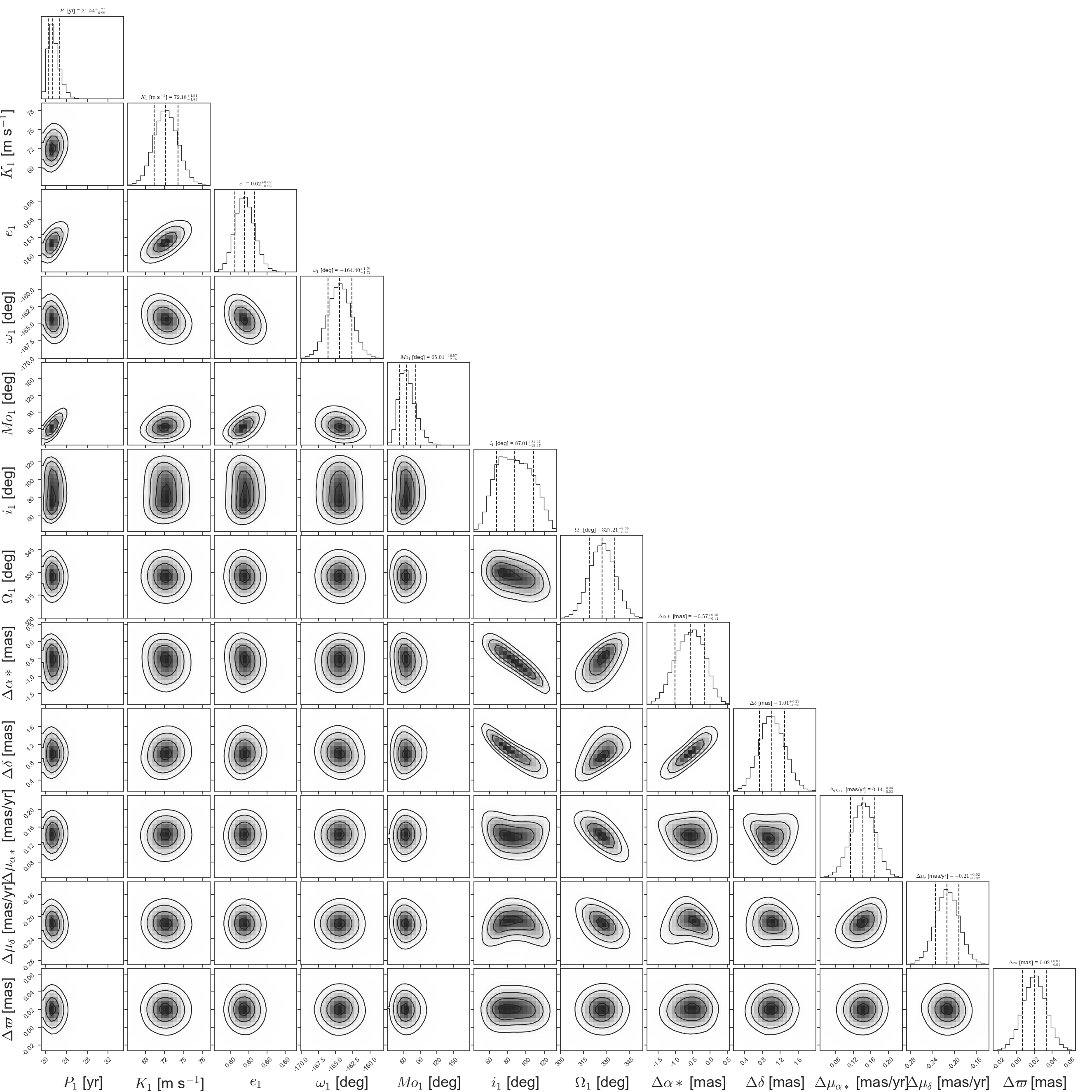}
    \caption{Similar to Figure~\ref{fig: 48265corner}. Corner plot of the posteriors for HD 68475 b.}
    \label{fig: 68475corner}
\end{figure}

\begin{figure}[htp]
    \centering
    \includegraphics[width=18cm]{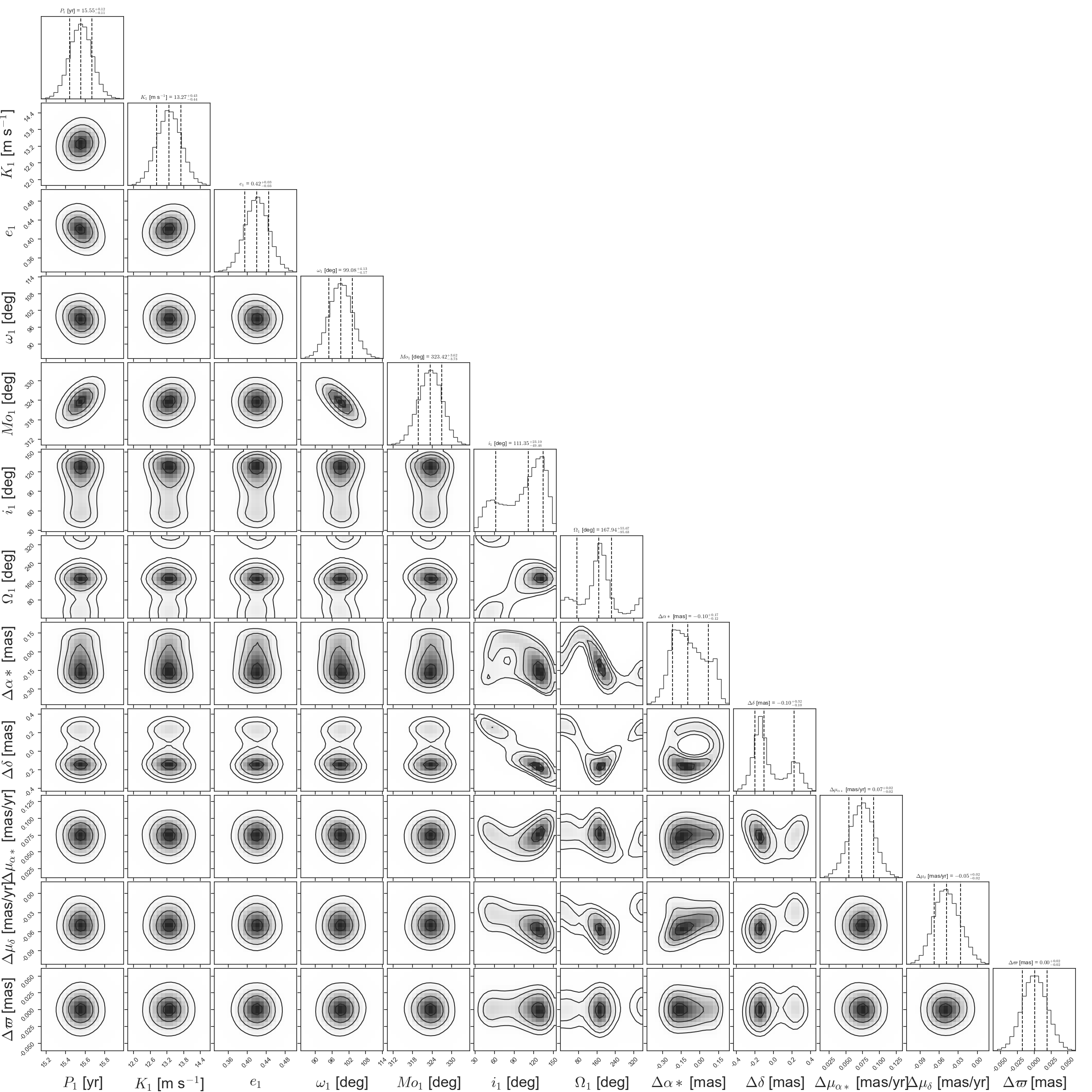}
    \caption{Similar to Figure~\ref{fig: 48265corner}. Corner plot of the posteriors for HD 100508 b.}
    \label{fig: 100508corner}
\end{figure}

\begin{figure}[htp]
    \centering
    \includegraphics[width=18cm]{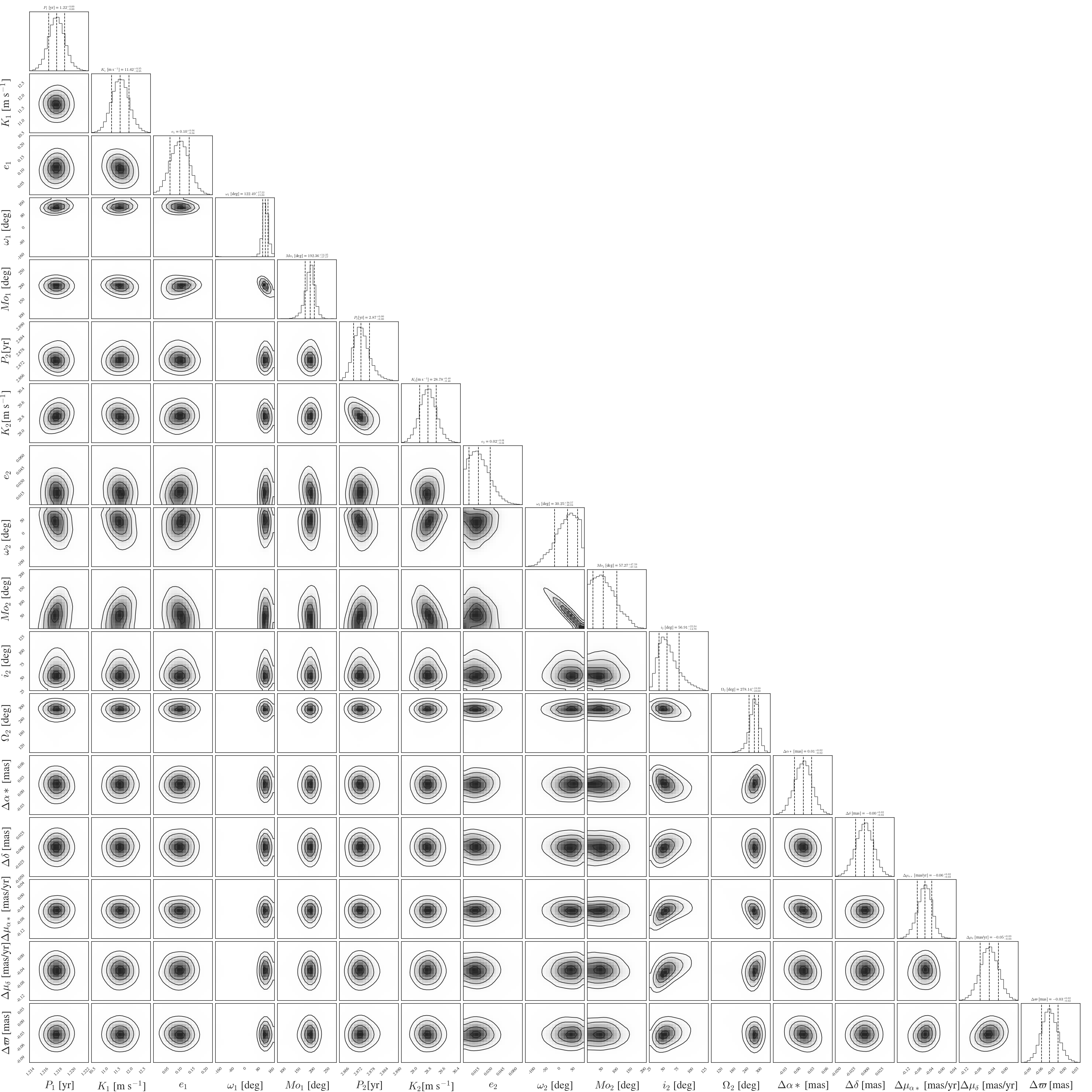}
    \caption{Similar to Figure~\ref{fig: 48265corner}. Corner plot of the posteriors for HD 114386 b and c.}
    \label{fig: 114386corner}
\end{figure}

\begin{figure*}[htp]
   \centering 
   \includegraphics[width=18cm]{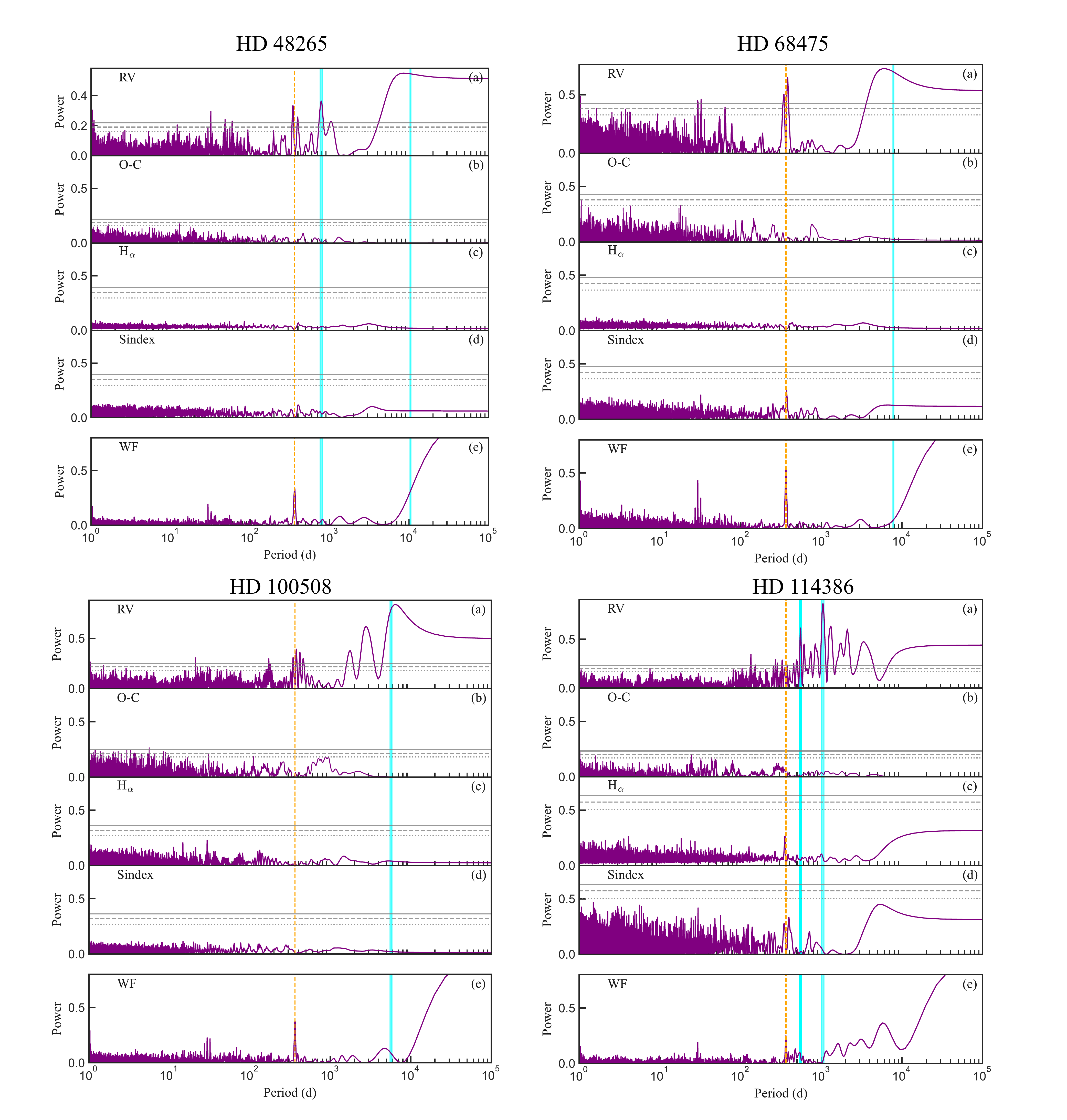}      \caption{Generalized Lomb–Scargle (GLS) periodograms for HD 48265, HD 68475, HD 100508, and HD 114386. For each target, panels show GLS power of the RVs (a), residuals (b), $H\alpha$ index (c), S-index (d), and window function (e). The blue shaded regions highlight strong periodic signals that coincide with the orbital periods of the reported planets, while the orange dashed line marks a signal near 365 days, which arises from the annual sampling of the data. The activity indicators ($H\alpha$ and S-index) show no significant power at the planetary periods, supporting that the observed RV variations are not caused by stellar activity.} 
       \label{fig: gls}
\end{figure*}

\begin{figure*}[htp]
   \centering 
   \includegraphics[width=12cm]{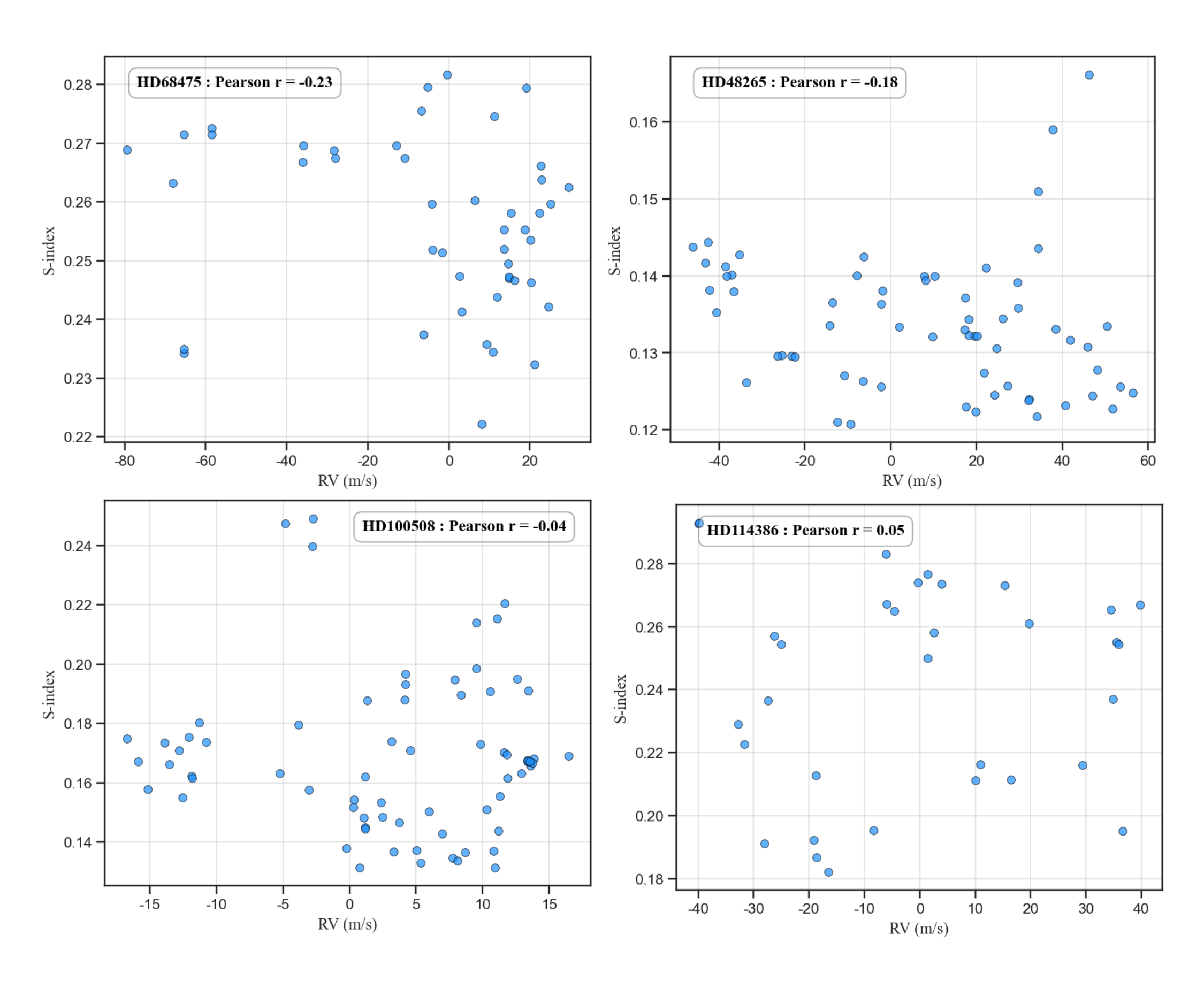}      \caption{Pearson correlation between RV and S-index for host star. The Pearson coefficient (r) is calculated to be $| r|< 0.3$, indicating a very weak correlation, suggesting that the RV signal is independent of stellar activity.} 
       \label{fig: pearson}
\end{figure*}

\begin{figure*}[htp]
   \centering 
   \includegraphics[width=16cm]{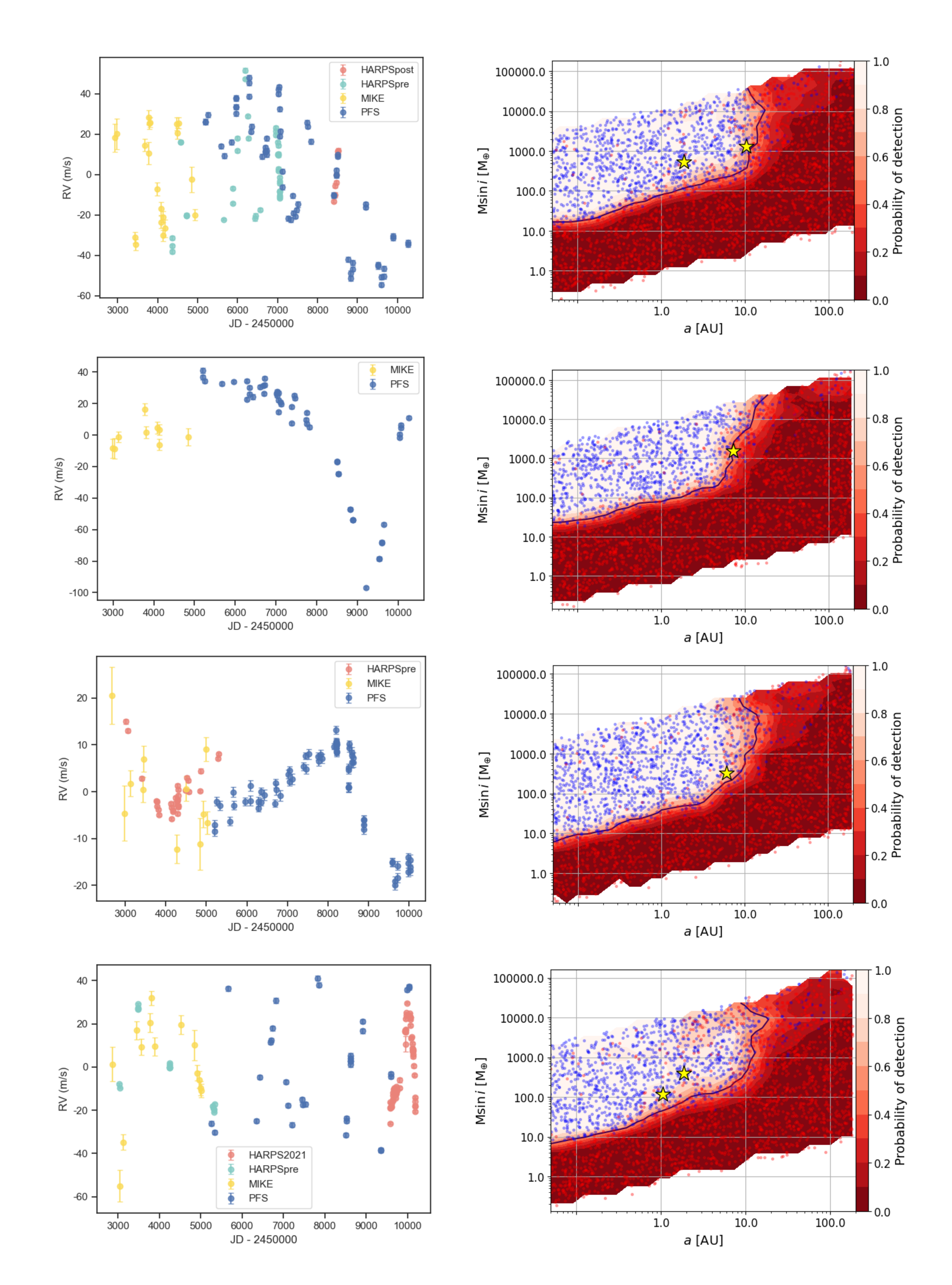}      \caption{Search completeness of our targets, assessed with injection–recovery tests using RVSearch \citep{2021ApJS..255....8R}. Left panels: RVs as a function of time. Right panels: injected signals in the $M\sin i$–$a$ plane. Blue dots: recovered injections; red dots: non-recoveries. Black contours: detection probability (solid line: 50\%). From top to bottom, we show results for HD 48265, HD 68475, HD 100508, and HD 114386.
} 
       \label{fig: comp}
\end{figure*}

\end{document}